\documentclass[12pt]{article}

\ifx\pdfoutput\undefined
\usepackage[dvips,bookmarks]{hyperref}
\else
\usepackage{hyperref}
\fi
\hypersetup{colorlinks=false,bookmarksopen,bookmarksnumbered,citecolor=blue,
   pdfstartview=FitH}

\usepackage[dvips]{graphicx}
\usepackage{latexsym}
\usepackage{color}
\usepackage{amssymb,amsfonts,amsmath}
\usepackage{graphicx} 
\usepackage{indentfirst}
\usepackage{enumerate}

\usepackage{bbm}
 
\usepackage{marvosym}

\parskip=\medskipamount

\def \un{\underline}
\newcommand {\cC}{{\cal C}}
\newcommand {\cD}{{\cal D}}

\newcommand {\cF}{{\cal F}}
\newcommand {\cG}{{\cal G}}

\newcommand {\cK}{{\cal K}}
\newcommand {\cL}{{\cal L}}
\newcommand {\cM}{{\cal M}}

\newcommand {\cN}{{\cal N}}
\newcommand {\cO}{{\cal O}}
\newcommand {\cP}{{\cal P}}

\newcommand {\cU}{{\cal U}}

\newcommand {\cW}{{\cal W}}

\def\a{\alpha}

\def\b{\beta}

\def\d{\delta}
\def\e{\epsilon}

\def\g{\gamma}
\def\G{\Gamma}

\def\l{\lambda}

\def\q{\theta}
\def\r{\rho}
\def\s{\sigma}
\def\t{\tau}

\def\z{\zeta}
\def\D{\Delta}
\def\F{\Phi}

\def\L{\Lambda}

\def\U{\Upsilon}
\def\X{\Xi}

\def\rd{{\rm d}}
\def\ri{{\rm i}}

\newcommand{\ve}{\varepsilon}                            

\newcommand{\pa}{\partial}                           
\newcommand{\hf}{\tfrac12}

\newcommand{\be}{\begin{equation}}
\newcommand{\ee}{\end{equation}}
\newcommand{\bea}{\begin{eqnarray}}
\newcommand{\eea}{\end{eqnarray}}
\newcommand{\non}{\nonumber}
\newcommand{\ba}{\begin{array}}
\newcommand{\ea}{\end{array}}

\newcommand{\1}{{\underline{1}}}
\newcommand{\2}{{\underline{2}}}

\newcommand{\bm}[1]{\mbox{\boldmath$#1$}}

\def\double #1{#1{\hbox{\kern-2pt $#1$}}}

\newcommand{\bsubeq}{\begin{subequations}}
\newcommand{\esubeq}{\end{subequations}}

\begin{document}

\begin{titlepage}

\begin{flushright}
PP-012-010\\
UUITP-09/12\\
June, 2012\\
\end{flushright}
\vspace{5mm}

\begin{center}
{\Large \bf Six-dimensional Supergravity and\\ Projective Superfields}\\ 
\end{center}

\begin{center}

{\large  
William D. Linch {\sc iii}~\,$\hspace{-6pt}{}^{\small \mbox\Pisces}$\
and 
Gabriele Tartaglino-Mazzucchelli~\,$\hspace{-6pt}{}^{\small \mbox\Virgo}$\
} \\
\vspace{5mm}

\footnotesize{
${}^{\small \mbox\Pisces}${\em Departamento de Ciencias F$\acute{\imath}$sicas,
Facultad de Ciencias Exactas,\\
Universidad Andres Bello,
Santiago de Chile\\
and\\
Center for String and Particle Theory, Department of Physics,\\
University of Maryland, College Park, MD 20472.}
~\\ 
~\\
${}^{\small \mbox\Virgo}${\em Theoretical Physics,
Department of Physics and Astronomy,\\
Uppsala University Box 516,
SE-751 20 Uppsala, Sweden\\ 
and\\
School of Physics M013, The University of Western Australia\\
35 Stirling Highway, Crawley W.A. 6009, Australia.\footnote{Corresponding address as of June, 2012.}}
}
~\\


\end{center}
\vspace{5mm}

\begin{abstract}
\baselineskip=14pt
We propose a superspace formulation of $\cN=(1,0)$ conformal
supergravity in six dimensions. 
The corresponding superspace constraints are invariant 
under super-Weyl transformations generated by a real scalar parameter. 
The known variant Weyl super-multiplet is recovered by coupling the geometry to a super-3-form tensor multiplet. Isotwistor variables are introduced and used to define projective superfields. 
We formulate a locally supersymmetric and super-Weyl invariant action principle in projective superspace. Some families of dynamical supergravity-matter systems are presented.
\end{abstract}

{\footnotesize
\begin{flushleft}
~\\
{${}^{\small \mbox\Pisces}$ \href{mailto:wdlinch3@gmail.com}{wdlinch3@gmail.com}}\\
{${}^{\small \mbox\Virgo}$ \href{mailto:gabriele.tartaglino-mazzucchelli@physics.uu.se}{gabriele.tartaglino-mazzucchelli@physics.uu.se}.
}
\end{flushleft}
}

\vfill
\end{titlepage}

\newpage
\renewcommand{\thefootnote}{\arabic{footnote}}
\setcounter{footnote}{0}

\tableofcontents{}
\vspace{.4cm}
\bigskip\hrule

\section{Introduction}

Recently, new off-shell superspace techniques have been developed to study
general 
supergravity-matter systems with eight real supercharges in various space-time dimensions.
These are based on the use of projective superspace,
introduced in the 1980s by
Karlhede, Lindstr\"om, and Ro\v cek to study rigid 4D, $\cN=2$ supersymmetry
\cite{KLR,LR}.\footnote{See also \cite{projective-revitalized} and references \cite{LRtwistor} and \cite{KuzReview} for reviews on flat 4D, $\cN=2$ projective superspace.}
Analogously to harmonic superspace \cite{harm1,harm}, projective superspace
is based on the extended superspace 
${\mathbb R}^{4|8}\times \mathbb{C}P^1$
where the projective coordinates are related to the 
SU(2) R-symmetry  group
of the extended supersymmetry algebra, 
an idea first introduced in the seminal paper \cite{Rosly}. 

The first attempt to extend projective supermultiplets to curved supersymmetry was undertaken in 2007
in a study of matter couplings in 5D, $\cN=1$ anti de-Sitter superspace \cite{KT-M}.
This was subsequently adapted to supergravity in various dimensions
in a series of papers, chronologically:
 5D in \cite{KT-M_5D,KT-M_5Dconf};
4D in \cite{KLRT-M_4D-1,KLRT-M_4D-2}; 2D in \cite{GTM_2D_SUGRA}; and
3D in \cite{KLT-M_3D_SUGRA}.
The formalism is based on two central ingredients: 
(i) a covariant geometric description in superspace of the supergravity multiplet;
(ii)
the existence of 
covariant projective multiplets, which are generalizations of 
 the superconformal projective multiplets introduced by Kuzenko in \cite{K,K2},
 and a locally supersymmetric and super-Weyl invariant 
action principle that is consistently defined on the curved
geometry of ingredient (i).

In many respects, the curved projective superspace formalism has proven to resemble
the covariant Wess-Zumino superspace approach to 4D, $\cN = 1$ supergravity \cite{WZ-s}.
However,
while a prepotential description of 4D, $\cN=2$ conformal supergravity
was given in harmonic superspace in \cite{Galperin:1987em},
its  relationship to standard geometrical methods of curved superspace has not 
yet been elaborated in detail.
A synthesis of curved harmonic and projective superspace could provide a coherent superspace description 
of 4D, $\cN = 2$ supergravity, along the lines of the Gates-Siegel 
approach to the 4D, $\cN = 1$ case \cite{SG}.
Besides the calculational advantages this affords ({\it e.g.}\,background field quantization), such an understanding has applications in covariant string theory.
These descriptions are (necessarily) closely related to the projective \cite{Berkovits:1999du,Linch:2006sh} and harmonic superspace formalisms \cite{Siegel:1995px,Siegel:2011sy}.
A particularly relevant example is that of the pure spinor formalism \cite{Berkovits:2000fe} compactified on a K3 surface, where the physical state conditions on the unintegrated vertex operators are automatically formulated in terms of analyticity conditions in 6D, $\mathcal N=(1,0)$ projective superspace \cite{Chandia:2011wd,Chandia:2011su}. Addition of the ``missing'' harmonics as non-minimal variables allows for a (simpler) description of the physical state conditions and the integrated vertex operators in harmonic superspace \cite{Chandia:2011su}.

This paper is devoted to the continuation of the aforementioned program and to the demonstration that 
a projective superspace formalism can be efficiently implemented also in the case of 
six space-time dimensions.
As a step toward the definition of six-dimensional curved projective multiplets, one first needs to 
indentify a proper geometric description in superspace of off-shell, $\cN=(1,0)$ supergravity.
In a standard fashion, a starting point to describe off-shell
supergravity systems is the coupling of the Weyl multiplet of conformal supergravity
 to matter compensators.
This is possible both in components, through the superconformal tensor calculus techniques 
(see \cite{Superconformal-tensor-calculus} for standard references),
and in superspace.
In components, the Weyl multiplet of 6D, $\cN=(1,0)$ conformal supergravity was constructed in
reference \cite{BergshoeffMZ}.
To our knowledge, however, a 
geometric description of the Weyl multiplet in six dimensions, 
analogous in spirit to the 4D, $\cN=2$ geometry of Howe \cite{Howe}, has hitherto not been 
fully developed.\footnote{
It is worth pointing out that 
a prepotential formulation of the Weyl multiplet
was given by Sokatchev in 6D harmonic superspace  \cite{SokatchevAA}.}

In this paper, we begin to fill this gap 
by presenting a superspace geometry suitable to the description of $\cN=(1,0)$ conformal
supergravity in six dimensions.
Specifically, our geometry naturally describes 
the 40+40 components of the Weyl supermultiplet of \cite{BergshoeffMZ} in superspace, in the 
form
having the ``matter'' components of the multiplet described by an anti-self-dual 3-form 
$W^{-}_{abc}$, a positive-chirality spinor $\chi^{\a i}$, and a real scalar $D$. 
We will refer to this Weyl multiplet as the type-{\sc i} multiplet.
In reference \cite{BergshoeffMZ}, it was shown that there is a second 40+40 Weyl multiplet
possessing as matter fields a scalar $\s$, 
a 2-form tensor $B_{ab}$, and a negative chirality spinor $\psi_{\a i}$;
we will refer to this as the type-{\sc ii} Weyl multiplet.
Such a formulation is engineered 
by coupling the type-{\sc i} multiplet to an on-shell tensor multiplet 
\cite{Howe:1983fr,Koller:1982cs}
and solving for the type-{\sc i} matter fields in terms of the fields of the tensor by using
the equations of motion of the latter. 
The same mechanism can be used to describe the type-{\sc ii} Weyl multiplet in superspace as we 
will show 
by coupling the type-{\sc i} superspace geometry
to a tensor multiplet described in terms of a closed super 3-form
(first introduced in the flat case in \cite{BergshoeffQM}).

Having constructed a superspace geometry suitable to the description of six-dimensional Weyl 
multiplets, the 
consistent definition of six-dimensional covariant projective superfields in this supergravity 
background
proceeds exactly as in the lower-dimensional cases.
In this paper, we will focus on such technical problems
as the construction of the 6D, $\cN=(1,0)$ multiplets, the
projective action principle, and the analytic projection operator.
We leave the applications of our results,
some of which we set out in the Conclusion (section \ref{Conclusion}),
for future investigation. Our hope is that the techniques we are starting to develop here will
be of use not only in extending classic results ({\it e.\,g.\,}\cite{NishinoDC,NishinoGK})
but also newer ones which have arisen 
in the resurgence of interest in 6D, $\cN=(1,0)$
 supersymmetry and supergravity;
see, for example, \cite{Metsaev:2010kp,Coomans:2011ih,Samtleben:2011fj,HD-6D,Akyol:2012cq}.

This work is organized into two parts with the supergeometrical part concentrated in section 
\ref{SuperGeometry} and the projective superspace part presented in section \ref{projSuper}. 
We begin the first part with the construction of the curved superspace geometry and give the 
dimension $\leq\frac32$ commutator algebra and torsion constraints in section \ref{SGeom}. In 
section \ref{SWeyl}, we give the super-Weyl transformations compatible with this geometry and 
use them to elucidate the relation to the type-{\sc i} multiplet of the superconformal tensor calculus. 
In section \ref{STypeII}, we solve the Bianchi identities of a closed super-3-form in the type-{\sc i} 
background and re-interpret the resulting supergravity-matter system in terms of the type-{\sc ii} 
Weyl multiplet. 
The second part begins with the construction of covariant projective superfields in six dimensions 
and the analytic projection operator in section \ref{cov-proj}. In section \ref{SAction}, we define the 
projective action principle, prove its consistency, and give families of examples of dynamical 
projective supergravity-matter systems. We conclude in section \ref{Conclusion} with some
reflection on our results and a description of future work
and open problems. Our 
conventions are defined in appendix \ref{appA} and the requisite properties of the analytic 
projection operator are demonstrated in appendix \ref{AnalyticProjectorAppendix}.

\section{6D, $\cN=(1,0)$ Supergravity in Superspace}
\label{SuperGeometry}

In this section, we present a new curved superspace geometry
suitable to the description of $\cN=(1,0)$ conformal supergravity in six dimensions.
In the spirit of Howe and Tucker \cite{Howe-Tucker}, we will see that 
the geometry is invariant 
under super-Weyl transformations generated by an unconstrained real scalar superfield. 
For this reason, the geometry will describe the 40+40 components of the 
type-{\sc i} Weyl multiplet and, once coupled to a tensor multiplet super 3-form, the type-{\sc ii} Weyl multiplet.
We refer the reader to the following list of references for previous work
on flat superspace and multiplets in six dimensions: \cite{Koller:1982cs,Howe:1983fr,Kugo:1982bn,oweAR,DragonNV,GrundbergLindstrom_6D,GatesPenatiTartaglino_6D}.
For the use of curved superspace to describe supergravity multiplets in six dimensions, see
\cite{BreitenlohnerRQ,GatesQV,GatesJV,SmithWAA,AwadaER,BergshoeffSU,BergshoeffRB}.

Our goal is to develop a formalism of differential geometry 
in a curved six-dimensional,  $\cN=(1,0)$
superspace $\cM^{6|8}$
that is locally parametrized by real   bosonic ($x^m$) 
and real fermionic ($\q^{\mu}_{i}$) coordinates
\bea
z^{M}=(x^m,\q^{\mu}_{i})~,
 \qquad
m=0,\cdots,3;5,6
~,~~~
\mu=1,2
~,~~~
i
={\1},\2~.
\eea
A natural condition on such a geometry is that it reduce to six-dimensional, $\cN=(1,0)$ Minkowski superspace in the flat limit .
Let us, to this end, recall that 
the  spinor covariant derivatives $D_\a^i$ 
associated with 6D, $\cN=(1,0)$ Minkowski superspace
satisfy the anti-commutation relations 
\bea
\{D_\a^i,D_\b^{j}\}=-2\ri \, \ve^{ij}(\g^c)_{\a\b}\,\pa_c~.
\eea
An explicit realization of  $D_\a^i$ is given by the expression
\bea
D_\a^i=\frac{\pa}{\pa\q^{\a}_{i}}
-\ri \,(\g^b)_{\a \b} \, \q^{\b i}\,\pa_{b}~.
\eea
Given a superfield $F$ of Grassmann parity $\e (F)$, the conjugation rule of its covariant derivative 
is
\bea
\overline{(D_\a^i F)}=-(-)^{\e(F)}D_{\a i} \bar{F}~,
\label{c.c.1}
\eea
with $\bar F:=(F)^*$ the complex conjugate of $F$.
Details of our notation and conventions are given in appendix \ref{appA}.

\subsection{{The Algebra of Covariant Derivatives}}
\label{SGeom}
For our curved geometry, we choose the structure group to be
$\mathrm{SO}(5,1)\times \mathrm{SU}(2)$.
The covariant derivative $(\cD_{{} A})= (\cD_{{} a} , \cD_{{} \alpha}^{ i})$ expands as
\begin{eqnarray}
\cD_{{} A} = E_{{} A} +\Omega_{{} A} +\Phi_{{} A}
~,
\end{eqnarray}
where
\begin{eqnarray}
E_{{} A} =E_{{} A}{}^{{} M}(z)\partial_{{} M}
~,~~~
\Omega_{{} A} = \frac12 \Omega_{{} A}{}^{{} b{} c}(z) M_{{} b {} c}
~,~~~
\Phi_{{} A}=  \Phi_{{} A}{}^{ij}(z) J_{ij}~,
\end{eqnarray}
denote the frame form, the spin connection, and the SU(2) connection, respectively.
Here,
$\pa_M=\pa/\pa z^M$,
$M_{ab}=-M_{ba}$ is the Lorentz generator and
$J^{ij}=J^{ji}$ is the SU(2) R-symmetry generator.
 These are defined by their action on the spinor covariant derivatives as
\begin{eqnarray}
\label{LorentzSpin}
[ M_{{} a {} b}, \mathcal D_{{} \gamma}^{ k} ] = -\frac12 ({\gamma}_{{} a{} b})_{{} \gamma}{}^{{} \delta} \mathcal D_{{} \delta}^{ k}~,
&&~~~~~~
[ J^{ij}, \mathcal D_{{} \gamma}^{ k} ] = \varepsilon^{k(i} \mathcal D_{{} \gamma}^{ j)}~.
\end{eqnarray}
It follows that
\begin{eqnarray}
[ M_{{} a {} b}, \mathcal D_{{} c} ] = 2 \eta_{{} c [ {} a} \mathcal D_{{} b]}~.
\end{eqnarray}

The supergravity gauge group is generated by local transformations of the form
\bea
\d_K\cD_A=[K,\cD_A]~~\mathrm{where}~~
K=K^C(z)\cD_C+\hf K^{cd}(z)M_{cd}+\hf K^{kl}(z)J_{kl}
~,
\label{SUGRA-gauge-group1}
\eea
with all the gauge parameters obeying natural reality conditions but otherwise arbitrary.
Given a tensor superfield $T(z)$, its transformation rule is
\bea
\d_KT=KT ~.
\label{SUGRA-gauge-group2}
\eea
The covariant derivatives satisfy the (anti)commutation relations
\begin{eqnarray}
\label{TRF}
[ \mathcal D_{{} A}, \mathcal D_{{} B} \} &=& T_{{} A{} B}{}^{{} C} \mathcal D_{{} C} +\frac12 R_{{} A {} B}{}^{{} c {} d} M_{{} c{} d} +F_{{} A {} B}{}^{kl} J_{kl}
~,
\end{eqnarray}
with $T_{AB}{}^C$ the torsion, $R_{AB}{}^{cd}$ the Lorentz curvature, and
$F_{AB}{}^{kl}$ the SU(2) R-symmetry field-strentgh.
These tensor fields are related to each other by 
the  Bianchi identities:
\bea
\sum_{[ABC)}{[}\cD_{{A}},{[}\cD_{{B}},\cD_{{C}}\}\} =0~.
\eea
To describe conformal supergravity, we impose 
conventional constraints on the torsion. In the six-dimensional case we are considering, they can be chosen to be
\bsubeq\begin{eqnarray}
\label{dim0}
{T_{{} \alpha}^{ i}{}_{\beta}^{  j}{}^{{} c}=- 2\ri\ve^{ij} (\gamma^{{} c})_{{} \alpha {} \beta}}~, 
	&& (\textrm{dimension-}0)\\
\label{dim1/2}
T_{{} \alpha}^{  i}{}_{\, {} \beta}^{  j}{}^{{} \gamma}_{  k} = 0 
~	,~  T_{{} \alpha}^{ i}{}_{ \, {} b}{}^{{} c}= 0~,
	&& (\textrm{dimension-}\tfrac12)\\
\label{dim1}	
T_{{} a\, {} b}{}^{{} c}= 0 
~	,~  T_{{} a}{}_{ \, {} \beta (j}{}^{{} \beta}_{ \, k)}= 0~.
	&& (\textrm{dimension-}1)
\end{eqnarray}
\esubeq
These constraints are similar to the four-dimensional, $\mathcal N=2$ superspace geometry of \cite{Howe}, formally identical to the five-dimensional conformal supergravity constraints of \cite{KT-M_5Dconf}, and 
closely related to 
the six dimensional off-shell geometry of reference \cite{BreitenlohnerRQ}.

With the constraints so introduced, the solution of the Bianchi identities can be shown to imply
that the torsion and curvature tensors in (\ref{TRF}) are expressed 
in terms of a small number of {mass-dimension-1 real} tensor superfields $C_{{} a {} b {} c}$ and $C_c^{ij}$, and their covariant 
derivatives.
The torsion component $C_c^{ij}=C_c^{ji}$ is an iso-triplet and $C_{abc}=W_{abc}+N_{abc}$ is a 3-form, which we split into anti-self-dual ($W$) and self-dual ($N$) parts. 

In terms of these basic torsions, the graded commutation 
relations of the covariant derivatives are given by
\bsubeq
\begin{eqnarray}
\hspace{-1cm}
\label{Algebra-1}
\{ \mathcal D_{{} \alpha i}, \mathcal D_{{} \beta j} \}&=& 
	2 \ri \varepsilon_{ij} (\gamma^{{} a})_{{} \alpha {} \beta}\mathcal D_{{} a}
	+2\ri C_{{} a \, ij} ({\gamma}^{{} a {} b {} c})_{{} \alpha {} \beta} M_{{} b {} c}
	+4\ri\varepsilon_{ij}W^{{} a{} b{} c} (\gamma_{{} a})_{{} \alpha {} \beta} M_{{} b{} c}\\
	&&
	+4\ri\varepsilon_{ij}N^{{} a{} b{} c} (\gamma_{{} a})_{{} \alpha {} \beta} M_{{} b{} c}
	-6\ri\varepsilon_{ij}C_{{} a}^{kl} (\gamma^{{} a})_{{} \alpha {} \beta} J_{kl} 
	- \tfrac{8\ri}3 N^{{} a {} b {} c} ({\gamma}_{{} a {} b {} c})_{{} \alpha {} \beta}  J_{ij}
	~,\cr\cr
\label{Algebra-2}
[\mathcal D_{{} \gamma k}, \mathcal D_{{} a} ] &=& 
	-C^{{} b}_{kl}({\gamma}_{{} a{} b})_{{} \gamma}{}^{{} \delta}  \mathcal D^l_{{} \delta} 
	+W_{{} a{} b{} c}({\gamma}^{{} b{} c})_{{} \gamma}{}^{{} \delta} \mathcal D_{{} \delta k}
	+N_{{} a{} b{} c}({\gamma}^{{} b{} c})_{{} \gamma}{}^{{} \delta} \mathcal D_{{} \delta k}\cr
	&&+ \left[-\tfrac \ri2(\gamma_a)_{\gamma \delta}T_{bc}{}^{\delta}_k +\ri(\gamma_{b})_{\gamma \delta} T_{ca}{}^\delta_k\right]M^{bc}\\ 
	&&+\left[ 
	-(\gamma_a)_{\gamma \delta} \mathcal C^{\delta\,ij}_{k} 
		+\d_k^i\left(4\cN_{a\,\g}{}^{j}-3\cC_{a\, \g}{}^{j} \right)
		+5 (\gamma_a)_{\gamma \delta} \,\delta^{i}_k \,\left( \mathcal C^{\delta j} -\tfrac13 \mathcal W^{\delta j}\right)
	\right]J_{ij}
	~,
\nonumber		
\end{eqnarray}
\esubeq
where {$T_{ab}{}^{\g}_k$, $\mathcal C^{\a}_i$, 
and $\mathcal W^{\a}_i$} are defined below (c.f.~eq.~\ref{vvstorsion}, \ref{calC}, and \ref{calW}, resp.).
The dimension-1 superfields $C_{a\, ij}$, $W^{\alpha \beta}:= \tfrac16 W_{abc}(\tilde \gamma^{abc})^{\alpha \beta}$, 
and $N_{\alpha \beta}:= \tfrac16 N_{abc}(\gamma^{abc})_{\alpha \beta}$ satisfy additional constraints which follow 
from the Bianchi identities.\footnote{We are grateful to D. Butter for discussions that led to the correction of some 
Bianchi Identities appearing in previous versions of this paper.}
To display the content of these constraints more clearly, we first define
their Lorentz- and isospin-irreducible components 
\bsubeq
\begin{eqnarray}
\label{calC}
\mathcal D_{\gamma k } C_{a\, ij} &{=:}& \mathcal C_{a\, \gamma k\, ij} + (\gamma_a)_{\gamma \delta}\mathcal C^{\delta}_{ijk}
	+\varepsilon_{k(i}\mathcal C_{a\, \gamma j)} + \varepsilon_{k(i}(\gamma_a)_{\gamma \delta}\mathcal C^{\delta}_{j)}
	\label{DC}~,\\ 
\mathcal D_{\gamma k} N_{\alpha \beta}&{=:}&\mathcal N_{\gamma k\, \alpha \beta}
	+\check{\mathcal N}_{\gamma k\, \alpha \beta}~,\\
\label{calW}
\mathcal D_{\gamma k} W^{\alpha \beta}&{=:}& \mathcal W_{\gamma k}{}^{\alpha \beta} + \delta_\gamma^{(\alpha}\mathcal W^{\beta)}_k.
\end{eqnarray}
\esubeq
Multiple isospin indices are fully symmetrized as are multiple Lorentz indices of the same height (except for the case $\check {\mathcal N}$, which has a part proportional to a $\gamma$-matrix; c.f.~eq.~\ref{solN}), Lorentz indices at different
 heights have had their traces removed, and fields with both vector and spinor indices are 
 $\gamma$-traceless.
These properties are reflected in their explicit forms as solutions to the Bianchi identities:
\begin{eqnarray}
\begin{array}{lcl}
\mathcal C_{a\, \gamma k\, ij}  = 0~, &~~~
	& \mathcal N_{\gamma k\, \alpha \beta }=0~,\\ 
\label{solN}
\mathcal C^\delta_{ijk}=-\tfrac 16 (\tilde \gamma^b)^{\delta \beta} \mathcal D_{\beta(k} C_{b\, ij)}
~,
	&&\check{\mathcal N}_{\gamma k \, \alpha \beta}=(\gamma^a)_{\gamma(\alpha}
	{\mathcal N}_{a\,\beta)k}~,
	~~~
	(\gamma^a)_{[\alpha\beta} {\mathcal N}_{a\,\gamma]k}=0~,\\ 
\mathcal C_{a\, \beta j}= \tfrac23\Pi_{a\, \beta}^{c\, \gamma} \mathcal D^i_\gamma C_{a\, ij}
~,
	&&\mathcal W_{\gamma k}{}^{\alpha \beta} = \mathcal D_{\gamma k} W^{\alpha \beta} 
	-\tfrac25 \delta_\gamma^{(\alpha} \mathcal D_{\delta k}  W^{\beta)\delta}~,\\ 
\mathcal C^{\gamma k}= -\tfrac19 \mathcal D_{\delta l} C^{\delta\gamma\, lk} ~,
	&&\mathcal W^{\alpha i} = \tfrac 25 \mathcal D^i_\beta  W^{\beta\alpha}.
\end{array}	
\end{eqnarray}
Here, $\Pi_{a\alpha}^{b\beta} = \delta_a^b\delta_\alpha^\beta +\tfrac16(\gamma_a\tilde \gamma^b)_\alpha{}^\beta$ is the projector onto the $\gamma$-traceless subspace ({\it i.e.}~$(\tilde \gamma^a)^{\gamma \alpha}\Pi_{a\alpha}^{b\beta}=0=\Pi_{a\alpha}^{b\beta}(\gamma_b)_{\beta\gamma}$). 

Finally, the dimension-$\tfrac32$ torsion is given in terms of the remaining fields as
\begin{eqnarray}
\label{vvstorsion}
T_{ab}{}^{\gamma}_{k}=
	\tfrac \ri2 (\gamma_{ab})_{\beta}{}^\delta \mathcal W_{\delta k}{}^{\beta \gamma}
	+\ri(\tilde{\g}_{[a})^{\g\d}\left[\cC_{b]\, \d}{}_{k} -\cN_{b]\,\d}{}_{k}\right]
	+\ri( \gamma_{ab})_\delta{}^{\gamma}\left[\mathcal C^{\delta}_{k}-\tfrac 16 \mathcal W^{\delta}_{k}\right]~.
\end{eqnarray}
With this, the dimension-1 and dimension-$\tfrac32$ commutators are completely specified. It has been verified that the Bianchi identities are satisfied up to and including dimension $\tfrac32$. Futher details of the geometry are not required for the purposes of this paper and will be expounded upon elsewhere. 

\subsection{{Super-Weyl Transformations and the Type-{\sc i} Weyl Multiplet}}
\label{SWeyl}

A short calculation shows that the constraints 
(\ref{dim0})|(\ref{dim1})
are invariant under {arbitrary} super-Weyl transformations defined by
\bsubeq
\begin{eqnarray}
\label{WeylTrans}
\delta \mathcal D_{\alpha i} &=& 
	\frac12 \sigma \mathcal D_{\alpha i}
	-2(\mathcal D_{\beta i} \sigma)M_\alpha{}^\beta 
	+4(\mathcal D_{\alpha}{}^{j}\sigma) J_{ij}
	~,\\
\label{WeylTransB}	
\delta \mathcal D_a &=& 
	\sigma \mathcal D_a 
	-\frac \ri2 (\mathcal D^k\sigma)\tilde \gamma_a\, \mathcal D_k 
	- (\mathcal D^b\sigma ) \, M_{ab}
	-\frac \ri8 (\mathcal D^i \tilde\gamma_a \mathcal D^j\sigma) \, J_{ij}
	~ ,
\end{eqnarray}
\esubeq
where the parameter $\s(z)$ is a real, unconstrained superfield. The components of the dimension-1 torsion are required to transform as
\bsubeq
\begin{eqnarray}
\label{WeylC}
\delta C_{a\, ij} &=& \sigma C_{a\, ij} +\frac{\ri}{8}(\mathcal D_{(i} \tilde \gamma_a \mathcal D_{j)}\sigma)
~,\\
\delta W_{abc} &=& \sigma W_{abc}
~,\\
\delta N_{abc} &=& \sigma N_{abc} -\frac {\ri}{32}  (\mathcal D^k \tilde \gamma_{abc} \mathcal D_k \sigma)
~.
\end{eqnarray}
\esubeq

The transformations of  {$C_{a \,ij}$} and $N_{abc}$ contain in-homogeneous terms which can be used to gauge away many of their components.
The {anti-self-dual} 3-form $W_{abc}$ transforms homogeneously 
{and represents a superspace generalization of the Weyl tensor.}
{It can be shown that, by properly choosing a Wess-Zumino gauge for our superspace 
geometry, the surviving physical components 
embedded in the geometry contain the SU(2) field-strength, the gravitino curl, 
an anti-self-dual auxiliary 3-form of mass dimension-1,
an auxiliary spinor of positive chirality of mass dimension-$\frac32$, 
the Weyl tensor, and a real auxiliary scalar field of mass-dimension-2. }
The resulting multiplet describes the $(40+40)$-component Weyl supermultiplet \cite{BergshoeffMZ}
\begin{align}
e_a{}^m, \psi_m{}^{\alpha i}, \Phi_a{}^{ij}, W^-_{abc}, \chi^{\alpha i}, D~,
\end{align}
to which we will refer as the type-{\sc i} multiplet.
Here, $e_a{}^m$ is the sechsbein, 
$\psi_m{}^{\alpha i}$ the gravitino,
 $\Phi_a{}^{ij}$ the SU(2) connection, and $W^-_{abc}$, $\chi^{\alpha i}$, $D$ are the ``matter'' 
 fields.
Here, the component gauge fields and the gravitino are related to the $\theta=0$ components of the supersechsbein and superconnections whereas
the matter fields of the Weyl multiplet 
arise in our geometry as components of the Weyl superfield:
$W^-_{abc}=W_{abc}|_{\q=0}$, $\chi^{\alpha i}=\cW^{\alpha i}|_{\q=0}$
and $D=\cD_{\a i}\cW^{\a i}|_{\q=0}$.
As originally defined \cite{BergshoeffMZ}, this Weyl multiplet contains an additional dilatation gauge field $b_m(x)$ but this degree of freedom is pure gauge 
and one can choose to work in the gauge in which it
 vanishes. Such a gauge arises naturally in the superspace geometry we have 
 introduced here. 
 This situation is similar to the 5D conformal supergravity in superspace described in
 \cite{KT-M_5Dconf} and to Grimm's formulation of 4D supergravity
 \cite{Grimm}, as explained in detail in \cite{KLRT-M_4D-2}. In these superspace treatments, as with our geometry, the $b_m$ field does not arise.

\subsection{{{The Tensor Multiplet and the Type-{\sc ii} Weyl Multiplet}}}
\label{STypeII}
There is a second formulation of the Weyl supermulitplet in which the anti-self-dual 3-form $W$, auxiliary 
positive chirality spinor $\chi$, and auxiliary scalar $D$,
 are replaced by a tensor multiplet consisting of
a propagating scalar $\sigma$, a gauge 2-form tensor $B$, and a negative chirality tensorino 
$\chi$ \cite{BergshoeffMZ}. 
This alternate formulation,
to which we will refer as the type-{\sc ii} Weyl multiplet,
plays an important role in six-dimensional supergravity since it is the one
that, within the superconformal tensor calculus approach,
 can be consistently used to construct actions for general matter-coupled supergravity systems. 
(See, for example, \cite{Coomans:2011ih} for a recent discussion of six-dimensional Poincar\'e 
supergravity obtained by coupling the type-{\sc ii} Weyl multiplet to a linear multiplet.)
In this subsection, following the same logic used in the component case, we work out the superspace version of the type-{\sc ii} formulation by 
coupling the type-{\sc i} formulation to a tensor multiplet \cite{Howe:1983fr,Koller:1982cs}.
In flat space, the tensor multiplet has been constructed as a closed 3-form in superspace in
\cite{BergshoeffQM}.
It is natural to formulate the consistent curved tensor multiplet constraints
extending such a construction to our
curved superspace geometry.
To this end, we must work out the mass
dimension $\leq3$ part of the 3-form Bianchi identities in the supergravity background. 

The super-3-form $H$ can be written in local coordinates as 
\begin{align}
H= \frac1{3!}\mathrm d Z^{M_3} \mathrm d Z^{M_2} \mathrm d Z^{M_1} H_{M_1 M_2 M_3} =  \frac1{3!}E^{A_3} E^{A_2} E^{A_1} H_{A_1 A_2 A_3} 
~.
\end{align}
This form is closed, $\mathrm d H=0$, iff its components satisfy the Bianchi identities
\begin{align}
 \frac1{3!}\mathcal D_{[B} H_{A_1 A_2 A_3)} -  \frac1{2!\cdot 2!} T_{[BA_1|}{}^C H_{C|A_2A_3)} = 0~.
\end{align}
The dimension-2 condition is consistent with {the constraint}
\begin{align}
\label{Hsss}
H_{\alpha i  \beta j \gamma k} = 0~,
\end{align}
provided that
\begin{align}
H_{\alpha i  \beta j c} = 2\ri \varepsilon_{ij} (\gamma_c)_{\alpha \beta}\Phi
~,
\end{align}
where $\Phi$ is an arbitrary real scalar superfield.
Next, the dimension-$ \frac52$ identity is solved by
\begin{align}
H_{\alpha i b c} = - (\gamma_{bc})_\alpha{}^\beta \mathcal D_{\beta i}\Phi
~.
\label{DF}
\end{align}
Finally, the dimension-3 identity gives
the expression for the 3-form 
$H_{abc}=H^+_{abc}+H^{-}_{abc}$
\bsubeq
\bea
\label{3formFS}
H^+_{abc}&=&\left(\frac \ri8 \mathcal D^k \tilde \gamma_{abc} \mathcal D_k  
	-16N_{abc}\right) \Phi~,
\\
H^-_{abc}&=& -16W_{abc} \Phi~,
\label{3formFS-2}
\eea
\esubeq
divided here into its self-dual and anti-self-dual parts.
Additionally, it implies that $\Phi$ satisfies
\begin{align}
\label{TensorConstraint2}
\mathcal D_{(i} \tilde \gamma^a \mathcal D_{j)} \Phi +16\ri C^a_{ij} \Phi =0
~.
\end{align}
This constraint is super-Weyl invariant iff $\Phi$ has scaling-dimension equal to 2, that is, 
$\delta \Phi = 2\sigma \Phi$. It is the curved-space analogue of the flat space constraint 
$D_\alpha^{(i}D_\beta^{j)}\Phi=0$ which describes the tensor multiplet consisting of a 
scalar\footnote{To follow the nomenclature normally used in the literature 
for the component fields of the tensor multiplet, we call the lowest component field $\s(x)$.
Context should serve to distinguish this component from the super-Weyl parameter superfield $\s(z)$.}
$\sigma(x) \sim \Phi(z)|_{\q=0}$, a tensorino 
$\psi_{\alpha i}(x)\sim D_{\alpha i} \Phi(z)|_{\q=0}$, 
and a self-dual 3-form field-strength 
$h^+_{abc}(x)\sim D^k \tilde \gamma_{abc}  D_{k} \Phi(z)|_{\q=0}$
\cite{BergshoeffQM}.
Indeed, using (\ref{DF})--(\ref{3formFS-2}), one derives the Bianchi identity
\bea
\frac1{3!}\mathcal D_{[a} H_{bcd]} 
-  \frac1{2!\cdot 2!} T_{[ab}{}^{\g k} H_{cd] \g k} = 0
~,
\eea
which implies that, up to spinorial torsion terms, the 3-form superfield 
$H_{abc}\sim\cD_{[a}B_{bc]}$ is locally exact.
Finally, we note that the constraint (\ref{TensorConstraint2}) puts the tensor multiplet on-shell. This 
is most easily checked by taking the flat-space limit, $D_\alpha^{(i}D_\beta^{j)}\Phi=0$, and 
showing that it implies, for example, $\pa^a\pa_a\Phi=0$.
In the supergravity case, the equations are covariantized by the supergravity fields which provide additional interaction terms. 

In fact, it can be shown that the constraint (\ref{TensorConstraint2}) is equivalent to the condition
\begin{align} 
\label{tensor1}
\mathcal D_{\alpha (i} V^{\beta}_{j)}-\frac14\delta_\alpha^\beta \mathcal D_{\gamma (i} V^\gamma_{j)}=0~,
\end{align}
on a spinor potential superfield $V^{\alpha i}$, provided we identify
\begin{align}
\label{tensor2}
\Phi = \mathcal D_{\alpha i} V^{\alpha i}~.
\end{align}
In flat superspace, this multiplet was first introduced by Sokatchev in \cite{SokatchevAA}.
It is straightforward to verify that the new constraint is invariant under super-Weyl transformations 
iff $V$ is a super-Weyl tensor of scaling dimesion-$\frac32$: $\d_\s V^{\a i}=\frac{3}{2}\s V^{\a i}$. 
Furthermore, it is invariant under a gauge 
transformation so that, in the components of this new multiplet, the superfield
 3-form field-strength $H_{abc}=H^+_{abc}+H^-_{abc}$ is replaced with the exterior derivative of 
 its gauge 2-form potential $B_{ab}\sim \mathcal D_i\tilde \gamma_{ab}V^i$ \cite{SokatchevAA}. 
 It is non-trivial that $\Phi$, as defined in (\ref{tensor2}), is a super-Weyl tensor of scaling 
 dimension-2.
 
As first pointed out in reference \cite{BergshoeffMZ}, {provided that the scalar component 
$\sigma(x) \sim\F|_{\q=0}$
is everywhere non-vanishing,}
the equations of motion can be solved for the components $\{W^-, \chi , D\}$ in terms of the 
components $\{\sigma, B, \psi\}$.
The result is a description in terms of
the components of the type-{\sc ii} {Weyl} multiplet \cite{BergshoeffMZ}
{\begin{align}
e_a{}^m, \psi_m{}^{\alpha i}, \Phi_a{}^{ij}, \sigma, \psi_{\alpha i}, B_{ab}
~.
\end{align}
This formulation can be 
interpreted as arising by taking the {($40+40$)}-component type-{\sc i} multiplet, 
coupling to the {$11+8$} fields $\{\sigma, B, \psi\}$, and then imposing the 
{$11+8$} degrees of freedom of the equations of motion as constraints. 
In this interpretation, the tensor supermultiplet does not add any degrees of freedom to the type-{\sc i} multiplet overall. 
In our superspace language, assuming the superfield $\F(z)\ne0$ is everywhere non-vanishing,
this is equivalent to solving for the dimension-1 torsion
superfields of the type-{\sc i} geometry in terms of the tensor multiplet 3-form superfields.

This suggests a second mechanism to remove the newly added tensor-multiplet degrees of 
freedom:
Whenever the scalar field in the superfield ${\Phi}$ is nowhere-vanishing on the body of the 
supermanifold,
 it is evidently possible to use the super-Weyl parameter to 
gauge $\Phi\to 1$.
Equation (\ref{TensorConstraint2}) then reduces to $C_{a}^{ ij}= 0$ and equations 
(\ref{Hsss})--(\ref{3formFS}) become
\begin{align}
H_{\alpha i  \beta j \gamma k} = 0~,~
H_{\alpha i  \beta j c} = 2\ri \varepsilon_{ij} (\gamma_c)_{\alpha \beta}~,~
H_{\alpha i b c} = 0~,~
H^+_{abc}=-16N_{abc}~,~
H^-_{abc}=-16W_{abc}~.
\end{align}
This super-Weyl gauge corresponds to further strengthening the second conventional constraint in 
equation (\ref{dim1}) by imposing $T_{a\, \beta (j}{}^\gamma_{k)}\to0$. (Equivalent observations were made already in reference \cite{BergshoeffRB}.)
The residual Weyl transformations are constrained by (\ref{WeylC}) to satisfy 
$\mathcal D_{\alpha (i} \mathcal D_{\beta j)}\sigma =0$.

It is interesting to note that in five dimensions there is
a mechanism similar to the one just described to formulate a variant Weyl multiplet.
In fact, by coupling the five-dimensional Weyl multiplet to an abelian vector multiplet constrained
to satisfy the curved Chern-Simons equation of motion, one can solve it for the auxiliary fields 
of the standard Weyl multiplet and end up  with the so-called dilaton-Weyl multiplet
\cite{Ohashi,Bergshoeff}. See reference \cite{KT-M_5Dconf} 
for a description of this mechanism in superspace.

We conclude this section by comparing the six-dimensional variant to the lower-dimensional cases.
In $\mathrm D=4$ and $5$, 
vector multiplets with eight supercharges are of primary importance for conformal supergravity
since they possess a scalar field as their lowest component. 
For this reason, off-shell vector multiplets
are the most natural conformal compensators in 4D, $\cN=2$ and 5D, $\cN=1$ supergravity
and, once coupled to the Weyl multiplet,
 give rise to the so-called minimal multiplets \cite{Howe,BS,Howe5Dsugra}.
In six dimensions, on the other hand,
the lowest component  of an off-shell vector multiplet is a positive-chirality Weyl spinor.
In superspace, the 6D off-shell vector multiplet is described by a dimension-$\frac{3}{2}$
superfield-strength $F^{\a }_i$ constrained by \cite{Siegel:1978yi}
\begin{align}
\label{VM}
\cD_{\a}^{(i}F^{\b j)}-\frac14\d_\a^\b\cD_{\g}^{(i}F^{\g j)}=0~~~~ \mathrm {and }~~~~
\cD_{\a}^{i}F^{\a}_i=0 
\end{align}
which, compared with the tensor multiplet constraint (\ref{tensor1}), is missing the scaling-
dimension-$2$ scalar superfield (\ref{tensor2}).\footnote{Note that the tensor multiplet defined by the constraint (\ref{tensor1}), which is the first of the two vector multiplet constraints in (\ref{VM}),
has on-shell physical fields while the vector multiplet is off-shell.
The $V^{\a i}$-multiplet includes the following physical fields:
A 2-form gauge potential $B^{ab}\sim (\g^{ab})^\a{}_{\b}D_{\a i}V^{\b i}|_{\q=0}$;
a scalar dilaton $\s\sim D_{\a i}V^{\a i}|_{\q=0}$;
and a fermion $\chi_{ \a i}\sim D_{\a i}D_\b^{j}V^{\b}_j|_{\q=0}$.
The vector multiplet, on the other hand, consists of the following physical fields:
A gaugino $\l^{\a i}\sim F^{\a i}|_{\q=0}$;
a gauge field strength $F^{ab}\sim (\g^{ab})^\a{}_{\b}D_{\a i}F^{\b i}|_{\q=0}$;
and an auxiliary iso-triplet $Y^{ij}\sim D_\a^{(i}F^{\a j)}$.
One can check that the components of the vector multiplet that are also components of the $V^{\a i}$ superfield are pure gauge in the latter case.
On the other hand, the physical fields of $V^{\a i}$ that are responsible for putting the multiplet on-shell are precisely those killed by the second constraint in (\ref{VM}).}
Due to the differences just mentioned, there is no direct analogue of the minimal multiplet in six 
dimensions.
In some respects, the 6D tensor multiplet closely mimics features of the lower-dimensional vector 
multiplets. 
It has a scalar that naturally plays the role of a dilaton but the crucial difference is that the 6D tensor multiplet is on-shell.

\section{Six-dimensional Curved Projective Superspace}
\setcounter{equation}{0}
\label{projSuper}

Covariant projective supermultiplets have been used recently to efficiently describe matter 
couplings in extended supergravity.
This was first done in five dimensions \cite{KT-M_5D,KT-M_5Dconf},
then applied to the four-dimensional case \cite{KLRT-M_4D-1,KLRT-M_4D-2},
and recently extended to two \cite{GTM_2D_SUGRA} 
and then three \cite{KLT-M_3D_SUGRA} dimensions.
In this section, we continue this program
by showing that the existence of covariant projective supermultiplets 
is consistent with the geometry of section \ref{SuperGeometry}.
(Projective superfields in flat 6D, $\cN=(1,0)$ Minkowski superspace were first 
introduced in \cite{GrundbergLindstrom_6D} and further studied in 
\cite{GatesPenatiTartaglino_6D}.)
We then conclude with a presentation of a locally supersymmetric and super-Weyl invariant action principle.

\subsection{6D, $\cN=(1,0)$ Covariant Projective Superfields}
\label{cov-proj}

In defining curved projective multiplets, we follow the same procedure 
that has been successfully developed in the $2\leq {\rm D}\leq 5$ supergravity cases
\cite{KT-M_5D,KT-M_5Dconf,KLRT-M_4D-1,KLRT-M_4D-2,GTM_2D_SUGRA,
KLT-M_3D_SUGRA}.
We start by introducing an {\em isotwistor} variable $v^i=(v^\1,v^\2)\in 
{\mathbb C}^2 \setminus  \{0\}$, 
defined to be inert under the action of the supergravity structure group: 
$[M_{ab},v^i]=[J_{kl}, v^i]=0$.
Using this isotwistor, we define the covariant derivatives
\bea
\cD_\a^{(1)}:=v_i\cD_\a^{i}~.
\eea
Note that the $\cD^{(1)}_\a$ derivative is homogeneous of degree one in $v^{i}$. 
Our curved superspace is then extended to
$\cM^{6|8}\times {\mathbb C}P^1$, with the isotwistor variable interpreted as providing homogeneous coordinates $[v^{\un 1}\!:\! v^{\un 2}]$ of the complex projective line.
Tensor superfields on this extension are called {\em isotwistor superfields}.
A weight-$n$ isotwistor superfield $U^{(n)}(z,v)$
 is defined to be holomorphic on an open domain of 
${\mathbb C}^2 \setminus  \{0\}$ with respect to 
the homogeneous coordinates $v^i$  
for ${\mathbb C}P^1$
and is characterized by the  conditions: 
\begin{enumerate}[(i)]
\item \label{i} It is  a homogeneous function of $v$ 
of degree $n$, that is,  
\bea
U^{(n)}(z,cv)&=&c^nU^{(n)}(z,v)~, 
\qquad c\in \mathbb{C}^*~,
\label{weight-n}
\eea
\item \label{ii} The supergravity gauge transformations act on $U^{(n)}$ 
as follows:
\bsubeq
\bea
\d_K U^{(n)} 
&=& \Big(K^{{C}} \cD_{{C}} +\hf K^{cd}M_{cd}+ K^{kl} J_{kl}\Big)U^{(n)} ~,  
\label{localU}
\\
J_{kl}U^{(n)}(v)&=&
-\frac{1}{(v,u)}\Big(v_{(k}v_{l)}\,u^i\frac{\pa}{\pa v^i}
-nv_{(k}u_{l)}\Big)
U^{(n)}(v)~,
\label{JU}
\eea 
\esubeq
where
\bea
(v,u):=v^iu_i
~,~~~~~~
\d_j^i=\frac{1}{ (v,u)}\big(v^{i}u_j- v_j u^{i} \big)
~.
\label{completeness}
\eea
\end{enumerate}
The auxiliary variable $u_i$ is constrained by $(v,u)\ne0$ but is otherwise completely arbitrary.
By definition, $U^{(n)}$ is a function only of $v$ and not $u$; the same should be true
for its variation.
Indeed, due to (\ref{weight-n}),
the superfield $(J_{kl}U^{(n)})$ can be seen to be independent 
of $u_i$ even though the transformations in (\ref{JU}) explicitly depend on it.

With the definitions (\ref{i}) and (\ref{ii}) assumed, the set of isotwistor superfields is closed under
products and the action of the $\cD_\a^{(1)}$ derivative.
More precisely, given weight-$m$ 
and weight-$n$ isotwistor superfields $U^{(m)}$ and $U^{(n)}$,
 the superfield $(U^{(m)}U^{(n)})$ is a 
weight-$(m+n)$ isotwistor superfield and the superfield $(\cD_\a^{(1)} U^{(n)})$
is a weight-$(n+1)$ isotwistor superfield.
Note that, as implicitly indicated in (\ref{localU}),
 general isotwistor superfields are not restricted to be Lorentz scalar. 
Ultimately, the use of the extra isotwistor variable should be interpreted as an efficient way to deal
with superfields that are (in general, infinite-dimensional) representations of the SU(2) group;
see \cite{KLRT-M_4D-1,KLRT-M_4D-2}  for more details.

The most important property of isotwistor superfields is that the anti-commutator 
of $\cD_\a^{(1)}$ covariant derivatives 
is zero when acting on a Lorentz scalar, isotwistor superfield $U^{(n)}$.
In fact, from (\ref{Algebra-1}), one obtains the anti-commutation relation
\bea
\{ \cD^{(1)}_\a, \cD^{(1)}_\b \}&=&
-8\ri C^{(2)}_{\g(\a} M_{\b)}{}^\g
-16\ri N_{\a \b} J^{(2)}
~,
\label{D1D1}
\eea
where we have defined
\bea
C^{(2)}_{\a\b}:=v_iv_jC^{ij}_{\a\b}
~~~\mathrm{and} ~~~
J^{(2)}:=v^iv^jJ_{ij}
~.
\eea
The SU(2) generators appear in the previous algebra only in the combination defined by 
$J^{(2)}$ which 
can easily be shown to vanish when acting on general isotwistor superfields $J^{(2)}U^{(n)}=0$. 
If one imposes that $U^{(n)}$ be a Lorentz scalar, then
\bea
\{ \cD^{(1)}_\a, \cD^{(1)}_\b \}U^{(n)}&=&0
~.
\label{DDU=0}
\eea
We define a weight-$n$, covariant projective superfield
$Q^{(n)}(z,v)$ to be an isotwistor superfield ({\it i.e.} satisfying (\ref{i}) and (\ref{ii}))
constrained by the analyticity  condition
\be
\cD_\a^{(1)}Q^{(n)}=0~.
\label{analyticity}
\ee  
The consistency of the previous constraint is guaranteed by equation (\ref{DDU=0}) 
which takes the form of an integrability condition for the 
analyticity constraint (\ref{analyticity}).

In conformal supergravity, the important issue is whether
the projective multiplets can be made to vary
consistently under the super-Weyl transformations. 
Suppose we are 
given a weight-$n$, projective superfield $Q^{(n)}$ 
that transforms homogeneously: $\d_\s Q^{(n)}\propto  \s Q^{(n)}$. 
Its transformation law is, then, 
uniquely fixed to be 
\bea
\d_\s Q^{(n)}=2n \s\, Q^{(n)}~,
\label{super-Weyl-Qn}
\eea
by imposing super-Weyl invariance of the constraint  (\ref{analyticity}).

Given a projective multiplet $Q^{(n)}$,
its complex conjugate 
is not covariantly analytic.
However, 
one can introduce a generalized analyticity-preserving 
conjugation $Q^{(n)} \to \breve{Q}^{(n)}$, defined as
\be
\left[{Q}^{(n)} (v)\right]\breve{~}\equiv \bar{Q}^{(n)}\big(
\overline{v} \to 
\breve{v}\big)~, 
\qquad \breve{v} = {\rm i}\, \s_2\, v~, 
\ee
with $\bar{Q}^{(n)}(\overline{v}) $ the complex conjugate of $Q^{(n)}(v)$.
It follows that $\breve{\breve{Q}}{}^{(n)}=(-1)^nQ^{(n)}$
so that real supermultiplets can be consistently defined when 
$n$ is even.
The superfield $\breve{Q}^{(n)}$ is called the {\em smile-conjugate} of 
${Q}^{(n)}$.
Note that, geometrically, this conjugation is a composition of complex conjugation 
and the antipodal map on $\mathbb{C}P^1$.
A fundamental property is that
\bea
\left[ {\cD^{(1)}_{ \a} Q^{(n)}} \right]\breve{~}=(-1)^{\e(Q^{(n)})}\, \cD^{(1)}_{\a}
 \breve{Q}{}^{(n)}~,
\eea
implying that the analytic constraint (\ref{analyticity}) is invariant under smile conjugation.

A simple class of 6D projective superfields
is defined as
$G^{(m)}(z,v)=v_{i_1}\cdots v_{i_m}G^{i_1\cdots i_m}(z)$.
These 
are constructed in terms of the completely symmetric isotensor
superfields
$G^{i_1\cdots i_m}(z)=G^{(i_1\cdots i_m)}(z)$
and describe regular holomorphic tensor fields on the complex projective space
$\mathbb{C}P^1$ parametrized by the homogeneous coordinates $v^i$.
Provided that the SU(2) transformation rule of 
$G^{i_1\cdots i_m}$ is the standard one
\bea
J_{kl}G^{i_1\cdots i_m}=\d_{(k}^{(i_1}G_{l)}{}^{i_2\cdots i_m)}
~,
\eea
the superfield $G^{(m)}$ satisfies all the conditions to be isotwistor.
Moreover, the analyticity condition
$\cD_\a^{(1)}G^{(m)}=0$
is equivalent to the following constraint on $G^{i_1\cdots i_m}$:
\bea
\cD_\a^{(j}G^{i_1\cdots i_m)}=0~.
\eea
This constraint is consistent with the super-Weyl transformation rule
$\d_\s G^{i_1\cdots i_m}=2m\s G^{i_1\cdots i_m}$.
When $m=2p$, one can further constrain $G^{(2p)}$  to be smile-real which
 is equivalent to the condition
$\overline{(G^{i_1\cdots i_{2p}})}=G_{i_1\cdots i_{2p}}$. 
This kind of multiplet is known in $4D$, $\cN=2$ supersymmetry literature as an {\em $\cO(2p)$-multiplet}.
It is a generalization of the well-known linear multiplet $\overline{G^{ij}}=G_{ij}$ that has $p=1$; 
for an incomplete list of references see
\cite{N=2tensor,SSW,ProjectiveSuperspace2,LR,G-RLRvUW,GIO}.
Note that when $m=1$, $G^{(1)}=v_iq^i$, the (necessarily complex) superfield
$q^i$ satisfies $\cD_\a^{(i}q^{j)}=0$ and
describes a six-dimensional extension of the 
Fayet-Sohnius hypermultiplet \cite{Fayet-Sohnius}. 
It is necessarily on-shell as a consequence of the impossibility of adding a central charge to the 6D, $\cN=(1,0)$ algebra.

Instead of the homogeneous coordinates $[v^{\un 1}\!:\! v^{\un 2}]$, it is often useful
to work with an inhomogeneous local complex variable $\z$ that is invariant 
under arbitrary  projective rescalings  $v^i \to c\, v^i $, with $ c\in \mathbb{C}^*$.
In such an approach, one should replace $Q^{(n)}(z,v)$ with a new superfield 
$Q^{[n]}(z,\z) \propto Q^{(n)}(z,v)$, where $Q^{[n]}(z,\z) $ is  holomorphic 
with respect to  $\z$. Its explicit definition depends on the supermultiplet under 
consideration.  
The space ${\mathbb C}P^1$ can 
naturally be covered by two open charts in which $\z$ can be defined, 
and the simplest choice is:
(i) the north chart characterized by $v^{\1}\neq 0$;
(ii) the south chart with  $v^{\2}\neq 0$.
In the projective superspace literature, the classification of multiplets normally proceeds
by restricting to the north chart and depends on the pole structure of Laurent expansion in $\z$.
Analogously to the curved cases in dimensions $2\leq {\rm D}\leq 5$ \cite{KT-M_5D,KT-M_5Dconf,KLRT-M_4D-1,KLRT-M_4D-2,GTM_2D_SUGRA,KLT-M_3D_SUGRA}, six-dimensional projective superfields 
generically
possess an infinite 
number of standard superfields.
As an example, consider off-shell charged hypermultiplets.
In projective superspace these have a natural description in terms of the so-called arctic superfield:
A {\em weight-$n$ polar multiplet} is described in terms of {\em arctic} superfields
 ${\U}^{(n)} (z,v) $, and their {\em antarctic} smile-conjugates
$ \breve{{\U}}^{(n)}(z,v)$.
By definition, ${\U}^{(n)}$ is a projective superfield that is well-defined in the whole north chart of ${\mathbb C}P^1$ (conversely $ \breve{{\U}}^{(n)}(z,v)$ is well-defined in the whole south 
chart).
In the north chart, ${\U}^{(n)}=(v^{\1})^n \U^{[n]}$ and 
$ \breve{{\U}}^{(n)}=(v^{\2})^n \breve \U^{[n]}=(v^{\1})^n\z^n \breve \U^{[n]}$ 
are represented as
\bea
\U^{[n]} (z,\zeta) =\sum_{k=0}^{\infty}\z^k {\U}_k(z) ~,~~
\breve \U^{[n]}(z,\zeta) =\sum_{k=0}^{\infty}\frac{(-1)^{k}}{\z^k}\bar{{\U}}_k(z)~,~~~~
\z:=\frac{v^{\2}}{v^\1}~,
~~~~~~
\label{arctic1}
\eea
in terms of an infinite sequence of ordinary superfields $\{ {\U}_k(z) \}_{k=0}^\infty$
and their complex conjugates.
The spinor covariant derivative $\cD^{(1)}_\a$ can be represented as 
\bea
\cD^{(1)}_{\a} = (v^\1) \cD^{[1]}_{\a}(\z)~, \qquad \cD^{[1]}_{\a} (\z)
&=& \cD^{\2}_\a - \z \cD^{\1}_\a  ~.
\eea
From this representation, and the representation of the arctic multiplet in the north chart 
(\ref{arctic1}), 
it follows that the analyticity condition (\ref{analyticity}) 
nontrivially relates the superfield coefficients $\U_k(z)$ in the series.

Another important example not of the polar type and mentioned later is the smile-real 
{\em tropical multiplet}. 
A weight-0, real, tropical superfield 
$V^{(0)}(z,v)=V^{[0]}(z,\z) = \sum_{k=-\infty}^\infty \zeta^k V_k(z)$ is required to 
be well-defined only on $\mathbb C^*$, that is, $\mathbb CP^1$ with both north and 
south poles removed. The reality condition $V^{(0)}=\breve{V}^{(0)}$ implies that $\overline{V_k}=(-1)^kV_{-k}$.
A special case of this is given by the product of a weight-0 arctic field and its 
smile-conjugate $V^{(0)}\sim \breve{\U}^{(0)}\U^{(0)}$.
A more detailed classification of 6D covariant projective superfields will be 
considered elsewhere. 
(See \cite{GrundbergLindstrom_6D,GatesPenatiTartaglino_6D} for a discussion in the flat case.
In particular, it is shown in \cite{GatesPenatiTartaglino_6D} how the flat six-dimensional vector multiplet is described in terms of
a prepotential tropical superfield.)

For applications, it is crucial that the analyticity constraint defining projective superfields
can be solved in terms of unconstrained isotwistor superfields and an {\em analytic projection operator}.
We introduce the fourth-order operator
\bea
\label{AnalyticProjector}
\D^{(4)}:=
\Big(
\cD^{(4)}
-\frac{5\ri}{6}C^{(2)}{}^{\g\d}\cD^{(2)}_{\g\d}
-5\ri\cC^{(3)}{}^{\g}\cD^{(1)}_{\g}
-\frac{\ri}{4}(\cD^{(2)}_{\g\d}C^{(2)}{}^{\g\d})
+3C^{(2)}{}^{\g\d}C^{(2)}_{\g\d}
\Big)
~,
\label{anapro}
\eea
where
\bea
\cD^{(4)}:=-\frac{1}{96}\ve^{\a\b\g\d}\cD^{(1)}_\a\cD^{(1)}_\b\cD^{(1)}_\g\cD^{(1)}_\d
~,~~~~~~
\cD^{(2)}_{\a\b}:=\cD^{(1)}_{[\a}\cD^{(1)}_{\b]}=-\cD^{(2)}_{\b\a}~,
\eea
and 
\bsubeq
\bea
&&\ve^{\a\b\g\d}C^{(2)}_{\g\d}=\ve^{\a\b\g\d}(\g^a)_{\g\d}C^{(2)}_a
=2(\tilde{\g}^a)^{\a\b}C^{(2)}_a
=2C^{(2)}{}^{\a\b}~,
\\
&&
\ve^{\a\b\g\d}(\cD^{(1)}_\b C^{(2)}_{\g\d})=-12\cC^{(3)}{}^\a=-12\cC^\a{}^{ijk}v_iv_jv_k~,~~~
\cC^{(3)}{}^\a:=-\frac{1}{12}\ve^{\a\b\g\d}(\cD^{(1)}_\b C^{(2)}_{\g\d})
~.~~~~~~~~~
\eea
\esubeq
The superfield $\cC^\a_{ijk}$ is the dimension-$\frac32$ torsion component defined in (\ref{DC}).
The crucial property of the analytic projection operator is that,  given an arbitrary
weight-$(n-4)$ isotwistor, Lorentz scalar superfield $U^{(n-4)}$,
the superfield $Q^{(n)} $ defined by
\bea
Q^{(n)} (z,v) := \D^{(4)} U^{(n-4)} (z,v) ~, 
\eea
is a weight-$n$ projective superfield:
\bea
\cD^{(1)}_{\a}Q^{(n)}=0~.
\eea
Moreover, both $Q^{(n)}$ and $U^{(n-4)}$ can be required to
transform homogeneously with respect to super-Weyl transformations 
in which case
the transformations are fixed to be
\be
\d_\s U^{(n-4)} = (2n-2) \s U^{(n-4)}~,~~~~~~
\d_\s Q^{(n)}=2n\,\s Q^{(n)}
~.
\label{s-Weyl-U}
\ee

It is worth noting that the analytic projection operator can be also used to build a 
weight-$4$ projective multiplet $\cP^{(4)}(z,v)$ from a scalar, $v$-independent superfield $P(z)$.
In fact, for any $P(z)$, the superfield $\cP^{(4)}(z,v):=\D^{(4)}P(z)$
is a weight-$4$ projective
superfield. 
Moreover, if one wants both $P$ and $\cP^{(4)}$ to transform homogeneously
under super-Weyl transformations then they have to satisfy:
 $\d_\s P=6\s P$ and $\d_\s \cP^{(4)}=8\s \cP^{(4)}$.
A derivation of these statements is given in Appendix \ref{AnalyticProjectorAppendix}.
We conclude this subsection by giving the analytic projection operator in an equivalent 
form:\footnote{It is instructive to compare the six-dimensional analytic projection operator 
with the five-dimensional one of \cite{KT-M_5D,KT-M_5Dconf}. 
There, the projector was presented in the gauge $C_{\hat{a}}^{ij}=0$ 
($\hat{a}=0,\cdots,4$ is the 5D vector index in the notation of \cite{KT-M_5D})
with only the 5D scalar torsion $S^{ij}$ appearing in the projector. 
With an appropriate dimensional truncation, one can see that the coefficients
in the 6D and 5D projectors match.}
\bea
\D^{(4)}&=&\ve^{\a\b\g\d}\Big(
-\frac{1}{96}\cD_\a^{(1)}\cD_\b^{(1)}\cD_\g^{(1)}\cD_\d^{(1)}
-\frac{5\ri}{12}\cD_\a^{(1)}C^{(2)}_{\b\g}\cD^{(1)}_{\d}
-\frac{\ri}{8}(\cD^{(2)}_{\a\b}C^{(2)}_{\g\d})
+\frac{3}{2}C^{(2)}_{\a\b}C^{(2)}_{\g\d}
\Big)
~.~~~~~~~~
\label{anapro-2}
\eea
This expression will be useful in the next subsection.

\subsection{The Action Principle}
\label{SAction}
In this subsection,
we give a projective superfield action principle invariant under the supergravity gauge 
group and super-Weyl transformations and such that, in the flat limit, it reduces to the 
six-dimensional action of \cite{GrundbergLindstrom_6D,GatesPenatiTartaglino_6D}.
The latter is an extension of the one originally introduced in four dimensions in  \cite{KLR} 
and reformulated in a projective-invariant form in \cite{S}.
The result is a simple generalization of the action principle in the curved projective superspaces
in $2\leq {\rm D}\leq 5$ 
\cite{KT-M_5D,KT-M_5Dconf,KLRT-M_4D-1,KLRT-M_4D-2,GTM_2D_SUGRA,
KLT-M_3D_SUGRA}.

Let $\cL^{(2)}$ be a real projective multiplet of weight-2.
We assume that $\cL^{(2)}$ possesses the super-Weyl transformation
\bea
\d_\s \cL^{(2)}=4\s \cL^{(2)}~,
\label{super-Weyl-L++}
\eea
which complies with the rule (\ref{super-Weyl-Qn}).
We also introduce a real isotwistor superfield $\Theta^{(-4)}$ such that
\bea
\d_\s \Theta^{(-4)}=
-2\s \Theta^{(-4)}~,~~~~~~
\D^{(4)}\Theta^{(-4)}=1~.
\label{C-density}
\eea

Associated with $\cL^{(2)}$ and $\Theta^{(-4)}$ is the following functional 
\bea
\label{ProjectiveAction}
S&=&
\frac{1}{2\pi} \oint_C (v, \rd v)
\int \rd^6 x \,{\rm d}^8\q\,E\, \Theta^{(-4)}\cL^{(2)}~, 
\qquad E^{-1}={\rm Ber}\, (E_{A}{}^{M})~.
\label{InvarAc}
\eea
This functional is  invariant under arbitrary re-scalings
$v^i(t)  \to c(t) \,v^i(t) $, 
$\forall c(t) \in {\mathbb C}^*$, 
where $t$ denotes the evolution parameter 
along the integration contour.
By using that under super-Weyl transformations,  $E$ transforms as
\bea
\d_\s E=-2\s E~
\eea
and the transformation properties (\ref{super-Weyl-L++})--(\ref{C-density}), we find that
the functional $S$ is super-Weyl invariant.
The action (\ref{InvarAc}) is also invariant under arbitrary local supergravity gauge 
transformations (\ref{SUGRA-gauge-group1}), (\ref{SUGRA-gauge-group2}) and (\ref{localU}). 
The invariance under general coordinate and Lorentz transformations is trivial
given that both $\Theta^{(-4)}$ and  $\cL^{(2)}$ are Lorentz scalars.
The invariance under the SU(2) transformations can be proved similarly to the 
$2\leq {\rm D}\leq 5$ cases: First, we note that
\bea
U^{(-2)}:=\Theta^{(-4)}\cL^{(2)}
\eea
is a isotwistor multiplet of weight $-2$.
Then, one verifies that
\bea
K^{ij} J_{ij} U^{(-2)} = -\pa^{(-2)}\Big( K^{(2)} U^{(-2)}\Big)~,
~~~~~~
\pa^{(-2)}:=\frac{1}{(v,u)}u^i\frac{\pa}{\pa v^i}
~.
\eea
Next, since $K^{(2)} U^{(-2)}$ has weight zero, it is easy to see that 
\bea
(v, \rd v )\, K^{ij} J_{ij}  \,U^{(-2)} = -{\rm d}t \,
\frac{{ \rm d}  }{{\rm d}t} \, \Big(K^{(2)}U^{(-2)}\Big)~, 
\eea
where, again, $t$ denotes the evolution parameter along the integration contour in 
(\ref{InvarAc}). Since the integration contour is closed, the SU(2)-part of 
the transformations of $U^{(-2)}$ (\ref{localU}) does not contribute to the variation of 
the action (\ref{InvarAc}). 

The isotwistor superfield $\Theta^{(-4)}$ 
is used to ensure the invariance of the action
under super-Weyl and SU(2) transformations.
An important point is that, in general, the supersymmetric action
 (\ref{InvarAc}) is independent of the explicit form of
 $\Theta^{(-4)}$, which is just an auxiliary constructive tool.
 To prove this, we need one observation about the 
 analytic projection operator $\D^{(4)}$ (\ref{anapro}) or (\ref{anapro-2}). 
Specifically, let us show that $\D^{(4)}$ 
is symmetric under integration-by-parts.
In the geometry of section \ref{SuperGeometry},
the rule for integration-by-parts is
\be
\int \rd^6 x \,{\rm d}^8\q\,E\, \cD_{A} \, V^{A} =0~, 
\ee
with $ V^{A}=(V^{a}, V^{\a}_i) $  an arbitrary superfield. 
Introducing arbitrary isotwistor superfields 
$\Psi^{(-n)}$  and $\F^{(n-6)}$, and by using the form of the analytic projection operator given in (\ref{anapro-2}),
we find the symmetry relation
\bea
\frac{1}{2\pi} 
\oint_C (v,\rd v)
\int \rd^6 x \,{\rm d}^8\q\,E\,\Big{\{}
\Psi^{(-n)}\D^{(4)}\F^{(n-6)}
-\F^{(n-6)}\D^{(4)}\Psi^{(-n)}
\Big\}=0
~.
\label{Symm-proj}
\eea
Now, let $\cU^{(-2)}$ be a real isotwistor prepotential for the Lagrangian $\cL^{(2)}$ 
in (\ref{InvarAc}):
\bea
\cL^{(2)}=\D^{(4)}\cU^{(-2)}~.
\eea
By using (\ref{Symm-proj}) and $\D^{(4)}\Theta^{(-4)}=1$, we can re-express the action (\ref{InvarAc})
in the form
\bea
S&=&
\frac{1}{2\pi} \oint_C (v, \rd v)
\int \rd^6 x \,{\rm d}^8\q\,E\, \cU^{(-2)}~.
\label{InvarAc3}
\eea
If the Lagrangian $\cL^{(2)}$, and hence $\cU^{(-2)}$, is independent of $\Theta^{(-4)}$
then (\ref{InvarAc3}) makes manifest the independence of (\ref{InvarAc})
on $\Theta^{(-4)}$.

We point out that there is a freedom in the choice of $\Theta^{(-4)}$.
For instance, given a real weight-$m$ isotwistor superfield $\G^{(m)}$, 
 $\Theta^{(-4)}$ may be defined as
\bea
\Theta^{(-4)}:=\frac{\G^{(m)}}{\D^{(4)}\G^{(m)}}~,~~~~~~
\d_\s\G^{(m)}=2(m+3)\s\G^{(m)}
~.
\eea
Additionally, one can consider a real Lorentz scalar
and SU(2) invariant superfield} $P=P(z)$ such that
\bea
\Theta^{(-4)}:=\frac{P}{\D^{(4)}P}
~;~~~~~~
\d_\s P=6\s P
~.
\eea
Note that the use of  $P$ is inequivalent to that of a general, weight-0 isotwistor superfield 
$\Gamma^{(0)}$ which may have non-trivial dependence on the projective parameter $\zeta$ and 
is, as such, not invariant under SU(2) transformations.

Let us take the flat limit of the action (\ref{InvarAc}).
This, up to total flat vector derivatives, can be written as 
\bea
S&=&
\frac{1}{2\pi} \oint_C (v, \rd v)
\int \rd^6 x\,{\rm d}^8\q \,\check\Theta^{(-4)}L^{(2)}
=
\frac{1}{2\pi} \oint_C (v,\rd v)
\int \rd^6 x \, D^{(-4)}D^{(4)}\check \Theta^{(-4)}L^{(2)} \Big|_{\q=0} \non \\
&=& \frac{1}{2\pi}  \oint_C (v,\rd v)
\int \rd^6 x \, D^{(-4)} L^{(2)}\Big|_{\q=0}~,
\label{InvarAc-flat}
\eea
with  $L^{(2)}$, $\check\Theta^{(-4)}$, and  $D^{(4)}$ the flat-superspace limit of the 
Lagrangian $\cL^{(2)}$, 
the density $\Theta^{(-4)}$, and the 
analytic projector $\D^{(4)}$ (\ref{anapro}), respectively. 
Here, we have also introduced the operator
\bea
D^{(-4)}:=-{1\over 96}\ve^{\a\b\g\d}
D^{(-1)}_{\a} D^{(-1)}_{\b}D^{(-1)}_{\g}D^{(-1)}_{\d}~, \qquad
D^{(-1)}_{ \a} := \frac{u_i}{(v,u)} D^i_{ \a}~.
\eea
The flat action is invariant 
under arbitrary projective transformations of the form:
\be
(u_i\,,\,v_i)~\to~(u_i\,,\, v_i )\,R~,~~~~~~R\,=\,
\left(\begin{array}{cc}a~&0\\ b~&c~\end{array}\right)\,\in\,{\rm GL(2,\mathbb{C})}~.
\label{projectiveGaugeVar}
\ee
As it is explicitly  independent of $u$, the same invariance holds for  the curved-superspace action 
(\ref{InvarAc}).
This invariance is a powerful tool in 
superspace theories with eight supercharges. For example, in 5D, $\cN=1$ 
\cite{KT-M_5D}
and 4D, $\cN=2$ \cite{KT-M-normal} supergravity it has been used to reduce the projective action 
principle to components.
Clearly, the same techniques can
 be used in the six-dimensional case to reduce the action 
(\ref{InvarAc}).

One can rewrite the contour integral in the north 
chart of $\mathbb{C}P^{1}$, $v^{\1}\neq 0$,
 in terms of the inhomogeneous complex variable $\z$
\bea
v^{i} =
v^{\1}\z^i ~,~~
\z^i=(1,\z)~, ~~ \z_i= \ve_{ij} \,\z^j=(-\z,1)~,
~~~~\z=\frac{v^{\2}}{v^{\1}} \in \mathbb C~.
\label{zeta}
\eea
The Lagrangian $L^{(2)}(z,v)$ in the north chart can be rewritten as
\bea
L^{(2)}(z,v):=\ri (v^{\1})^2\z L(z,\z)
~.
\eea
Since the action and the Lagrangian are independent of $u_i$, we can make the conventional 
choice
\be
u_i =(1,0) ~, \qquad   \quad ~u^{i}=\ve^{ij }\,u_j=(0,-1)
~.
\label{fix-u-}
\ee   
The action (\ref{InvarAc-flat}) is, then, rewritten as
\bea
S&=&   \oint_C \frac{\rd\z}{2\pi\ri}
\int \rd^6 x \,\z\, (D^{\1})^4 L\Big|_{\q=0}~,~~~~
(D^{\1})^4:=-{1\over 96}\ve^{\a\b\g\d}
D^{\1}_{\a} D^{\1}_{\b}D^{\1}_{\g}D^{\1}_{\d}
~.
\eea
This expression is the rigid supersymmetric action in the 6D, $\cN=(1,0)$ projective superspace
of \cite{GrundbergLindstrom_6D,GatesPenatiTartaglino_6D}. 
Thus, our curved projective action principle is, as expected, a generalization of the
known flat one.

\subsection{Some Matter Systems}
We conclude this section with examples of supergravity-matter systems.
We start by considering two classes of projective superfield conformal compensators:
 an $\cO(2)$ multiplet, given by the real, linear superfield $G^{(2)}:=G^{ij}v_iv_j$
 and a weight-1, arctic multiplet $\U^{(1)}$ and its smile conjugate $\breve{\U}^{(1)}$ that describes the off-shell, charged hypermultiplet.

 Note that to use the linear multiplet as a proper compensator,
 $G^{ij}$  should be  nowhere-vanishing which is equivalent to $G:=\sqrt{G^{ij}G_{ij}}\ne0$.
This composite scalar and SU(2) invariant superfield, which has scale dimension 4,
 $\d G=4\s G$, can be used to choose the super-Weyl gauge $G=1$.
In this gauge, $\cD_\a^i G=\cD_\a^i 1=0$ which, 
together with the analyticity constraint 
$\cD_\a^{(i}G^{jk)}=0$, implies
 that $G^{ij}=w^{ij}$ is covariantly constant $\cD_{\a}^{i}w^{jk}=0$ wherefore
 also the SU(2) group is broken to the U(1) subgroup that leaves $w^{ij}$ invariant.
 By imposing the consistency of the supergravity algebra with the covariant constancy of $w^{ij}$,
 $\{\cD_\a^{i},\cD_\b^{j}\}w^{kl}=0$, one can easily see that, in this gauge, the
dimension-1  torsions satisfy\footnote{Similar gauges in superspace were used before
in 4D in \cite{KT-M-normal,Butter:2010jm} and in 3D in \cite{KLT-M_3D_SUGRA}.}
\bea
N_{abc}=0~,~~~~~~
C_a^{ij}=C_a w^{ij}
~.
\eea

The Lagrangian for the $\cO(2)$ multiplet compensator is given by
 \bea
\cL^{(2)}_{\rm SG-linear} = 
-G^{(2)} \ln \frac{G^{(2)} }{\ri \breve\Upsilon^{(1)}\Upsilon^{(1)} } 
~.
\label{O2sugra}
\eea
It encodes the dynamics of a massless improved linear multiplet coupled to conformal supergravity. It has the same form as the 4D, $\cN=2$ counterpart
given in \cite{Kuzenko:2008qz} as a locally-supersymmetric extension of the projective-superspace 
formulation \cite{KLR} for the 4D, $\cN=2$ improved tensor multiplet \cite{deWPV,LR83}.
The action (\ref{O2sugra}) is independent of the (ant-)arctic superfields $\U^{(1)},\breve{\U}^{(1)}$
which turn out to be pure-gauge superfields.
The Lagrangian (\ref{O2sugra}) can be shown to be dual to the Lagrangian for 
an arctic compensator coupled to conformal supergravity:
\bea
\cL^{(2)}_{\rm SG-hyper} = - \ri \,
\Upsilon^{(1)} \breve\Upsilon^{(1)}
~.
\eea
The duality map is the same as in reference \cite{Kuzenko:2008qz}.

By using the compensators, we can couple supergravity to general matter. We consider a few 
examples which are familiar from the lower-dimensional cases; we refer the reader to 
\cite{KLRT-M_4D-1,Kuzenko:2008qz} for the 
geometric interpretation of the models that follow.

Consider a system of interacting weight-1
 arctic  multiplets $\U^{(1) I} (v) $ and their smile-conjugates $ \breve{\U}^{(1)\bar I }(v)$ described by a Lagrangian of the form \cite{K2}
\bea
\cL_{{\rm NLSM-conf}}^{(2)} 
= {\rm i} \, K (\U^{(1)I}, \breve{\U}^{(1) \bar J})~.
\label{conformal-sm}
\eea
Here, $K(\F^I, {\bar \F}^{\bar J}) $ is a real function
of $n$ complex variables $\F^I$, with $I=1,\dots, n$, 
that satisfies the homogeneity condition
\bea
\F^I \frac{\pa}{\pa \F^I} K(\F, \bar \F) =  K( \F,   \bar \F)~.
\label{Kkahler22}
\eea
This Lagrangian represents a conformal non-linear sigma-model as in \cite{K2}.

Given a system of $m$ weight-0 arctic multiplets $\X^{i}$, 
$i=1,\cdots,m$,
and the conformal compensator $\U^{(1)}$, one can write
the Lagrangian 
\bea
\cL_{\rm NLSM-hyper}^{(2)}= \U^{(1)} \breve{\U}^{(1)} \,
\exp \Big\{ \cK (\X^{ i}, \breve{\X}^{\bar j}) \Big\}~.
\eea
The real function $\cK (\X^i, {\breve \X}^{\bar j})$ can be interpreted as a K\"ahler potential
since the Lagrangian is invariant under the transformation
\bea
{\U}^{(1)} ~\mapsto
~{\rm e}^{-\L(\X)} \, \U^{(1)}~, \qquad 
\cK(\X, \breve{\X} ) ~\mapsto
~ \cK(\X, \breve{\X} )+ \L(\X) + {\bar \L}( \breve{\X})~,
\eea
with $\L $ a holomorphic function.
In the dual picture, where the compensator is given by a linear superfield, the previous Lagrangian
is equivalent to
\bea
\cL_{\rm NLSM-linear}^{(2)}= G^{(2)}\cK (\X^{ i}, \breve{\X}^{\bar j})~.
\eea

Next, we consider a system of $n$ linear $\cO(2)$ multiplets $G^{(2)}_I$,
  $I=1,\dots, n$, coupled to conformal supergravity.
The Lagrangian takes form
\bea
\cL_{\rm SM-linear}^{(2)} = \cL (G^{(2)}_I)~,
\eea
where $\cL$ is a real homogeneous function of degree-$1$:
\bea
G^{(2)}_I \frac{\pa }{\pa G^{(2)}_I} \cL =\cL~.
\eea
More generally, it is possible to couple linear $\cO(2)$ multiplets and hypermultiplets 
in an arbitrary way
provided that the Lagrangian $\cL^{(2)}(G^{(2)},\U^{(1)},\breve{\U}^{(1)},\X,\breve{\X})$
is weight-2 in the sense that $\cL^{(2)}(c^2G^{(2)},c\U^{(1)},c\breve{\U}^{(1)},\X,\breve{\X})=
c^2\cL^{(2)}(G^{(2)},\U^{(1)},\breve{\U}^{(1)},\X,\breve{\X})$ with $c\in\mathbb{C}^*$.

We conclude by considering some composite, weight-2, scaling-dimension-4, real projective 
superfields built from tensors and vector field-strengths.
We begin by taking two tensor multiplets in the representations $\F$ and $V^{\a}_i$, 
introduced in section \ref{STypeII},
and coupling them through the composite $\cO(2)$ superfield
\begin{align}
\label{TT}
\cG^{(2)} : = \ri
(\mathcal D^{(1)}_\alpha \Phi) V^{\alpha {(1)}}
+ \frac\ri4 \Phi \mathcal D^{(1)}_\alpha V^{\alpha {(1)}}
~,~~~~~~~~~
V^{\alpha(1)}:=v_iV^{\alpha i}
~.
\end{align}
That this combination is analytic follows from a short calculation:
\begin{eqnarray}
\mathcal D^{(1)}_\beta \cG^{(2)} &=& 
	\ri(\mathcal D^{(1)}_\beta \mathcal D^{(1)}_\alpha \Phi) V^{\alpha {(1)}}
	-\ri(\mathcal D^{(1)}_\alpha \Phi) \mathcal D^{(1)}_\beta V^{\alpha {(1)}}
	+ \frac\ri4(\mathcal D^{(1)}_\beta \Phi) \mathcal D^{(1)}_\alpha V^{\alpha {(1)}}
	+ \frac\ri4 \Phi\mathcal D^{(1)}_\beta \mathcal D^{(1)}_\alpha V^{\alpha {(1)}}\cr
&=& 4C^{(2)}_{\beta \alpha} \Phi V^{\alpha {(1)}} 	+ \tfrac\ri4 \Phi\mathcal D^{(1)}_\beta \mathcal D^{(1)}_\alpha V^{\alpha {(1)}}=0.
\end{eqnarray}
Here, we are using the constraints (\ref{TensorConstraint2}) and (\ref{tensor1}) in the second equality. The third equality uses $\mathcal D^{(1)}_\beta \mathcal D^{(1)}_{\alpha} V^{\alpha {(1)}} = 16\ri C^{(2)}_{\beta \alpha }V^{\alpha{(1)}}$, which follows from the tensor constraint (\ref{tensor1}). 
Additionally, it is non-trivial but easy to check that this composite field is a super-Weyl tensor of scaling dimension 4: $\d\cG^{(2)}=4\s\cG^{(2)}$.
Of course, all the previous arguments also hold in the case that the two tensor multiplets are not 
independent one of one another but satisfy $\F=\cD_{\a i}V^{\a i}$ 
as in (\ref{tensor2}).

Comparison of the constraints to those of the vector multiplet (\ref{VM}) shows that the same arguments work if we formally replace the tensor potential $V^\alpha_i\to F^\alpha_i$ with the vector field-strength. 
Thus, the coupling of a vector and a tensor multiplet naturally gives rise to the weight-2
projective composite superfield \cite{BergshoeffMZ}
\begin{align}
\cF^{(2)} : =\ri (\mathcal D^{(1)}_\alpha \Phi) F^{\alpha {(1)}}+ \frac\ri4 \Phi \mathcal D^{(1)}_\alpha F^{\alpha {(1)}}.
\end{align}
If one, furthermore, considers a vector multiplet prepotential,
which can be shown to
be described by a weight-0, real, tropical superfield ${\bm V}:=V^{(0)}$,
then it is possible to construct the Lagrangian
\bea
\cL^{(2)}={\bm V}\cF^{(2)}
~.
\eea
This should be compared with the five-dimensional vector multiplet Lagrangian coupled to supergravity
\cite{KT-M_5D,KT-M_5Dconf}.

Finally, we point out that we can further extend the construction of the previous bilinear:
Consider a real weight-0 isotwistor superfield ${\bm \F}^{(0)}(z,v)$ 
and a real weight-1 isotwistor superfield ${\bm V}^{\a (1)}$ constrained by
\bsubeq
\bea
&&\big((\tilde{\g}_a)^{\a\b}\cD_\a^{(1)}\cD_\b^{(1)}
+16\ri C_a^{(2)}\big)
{\bm \F}^{(0)}(z,v)=0~,~~~~~~
\d_\s{\bm \F}^{(0)}=2\s {\bm \F}^{(0)}~,
\\
&&
\cD_\a^{(1)}{\bm V}^{\b (1)}-\frac{1}{4}\d_\a^{\b}\cD_\a^{(1)}{\bm V}^{\a (1)}=0~,~~~~~~
\d_\s{\bm V}^{\a(1)}=\frac{3}{2}\s {\bm V}^{\a(1)}~.
\eea
\esubeq
Then, analogously to the previous cases, the composite superfield
\begin{align}
{\bm \cL}^{(2)} : = \ri
(\mathcal D^{(1)}_\alpha {\bm \F}^{(0)}) {\bm V}^{\alpha {(1)}}
+ \frac\ri4 {\bm \F}^{(0)} \mathcal D^{(1)}_\alpha {\bm V}^{\a(1)}
~,
\label{crazy}
\end{align}
is a real, weight-2 projective superfield
such that $\d_\s{\bm \cL}^{(2)} =4\s{\bm \cL}^{(2)}$. Note that, in this case, 
${\bm \cL}^{(2)}$ is not an $\cO(2)$ multiplet.

The Lagrangian (\ref{crazy}) appears to be the projective superspace analogue of the harmonic superspace
Lagrangian introduced by Sokatchev to describe an off-shell tensor multiplet
\cite{SokatchevAA}. The latter was constructed by first taking 
a tensor multiplet of the $\Phi$-type and an independent tensor multiplet of the $V^{\alpha i}$-type
and lifting them to harmonic superspace by allowing them arbitrary dependence on the harmonics.
The construction of the projective action (\ref{crazy}) is analogous: We started with two copies of the tensor multiplet (in different representations) and took them off-shell by allowing them to have arbitrary dependence on the isotwistor variable $v^i$.

\section{Conclusion}
\setcounter{equation}{0}
\label{Conclusion}
In this paper, we have initiated the study of six-dimensional, $\cN=(1,0)$ supergravity in projective 
superspace. 
Beginning with the conventional constraints (\ref{dim0})--(\ref{dim1}),
 we provided the solution (\ref{Algebra-1})--(\ref{vvstorsion}) of the Bianchi identities up to and 
 including dimension-$\frac32$. Super-Weyl transformations (\ref{WeylTrans}, \ref{WeylTransB}) preserving this 
 geometry were computed and used to recover the components of the type-{\sc i} Weyl multiplet
of 6D, $\cN=(1,0)$ conformal supergravity. 
Coupling this multiplet to a closed super-3-form, we recovered the type-{\sc ii} Weyl multiplet.
 With the supergeometry understood, projective isotwistor variables were introduced and 
 used to define projective superfields. The defining constraint of such fields was solved by 
 constructing the analytic projection operator (\ref{AnalyticProjector}), which was subsequently 
 used to define a projective superspace action principle (\ref{ProjectiveAction}). This was checked 
 to be invariant under super-Weyl, local super-Poincar\'e, and local SU(2) transformations and reduced 
 to its flat limit which agrees with the flat actions of 
 \cite{GrundbergLindstrom_6D,GatesPenatiTartaglino_6D}.
 We concluded with the presentation of families of examples of such action 
 principles for supergravity-matter systems.

Clearly, much remains to be done to complete our understanding of 
$\cN=(1,0)$
supergravity in six-dimensional projective superspace. Perhaps the most pressing open problem 
is the construction of the projective superspace analogue of Sokatchev's harmonic supergravity 
\cite{SokatchevAA}. This construction exploits a remarkable combination of harmonic superspace 
prepotentials, both representations of the tensor multiplet, and the dynamical equations of the 
``matter'' components of the type-{\sc i} multiplet to avoid multiplet doubling. 
The Lagrangian (\ref{crazy}) is similar to the doubled-tensor compensator Lagrangian central to that 
construction. It would be of interest to confirm their equivalence and work out the detailed relation between these constructions.

Additional directions of study include
compactification to five dimensions and comparison with the work of 
\cite{KT-M_5D,KT-M_5Dconf}
and the recovery of our geometry from the gauging of the six-dimensional superconformal group 
along the lines of references \cite{Butter-1,Butter-2} 
which develope a direct link between superspace and superconformal tensor calculus.\footnote{An early attempt in six dimensions was made in reference
\cite{BreitenlohnerRQ}.}
More straightforward work in need of completion includes: 
the presentation of the  complete solution of the Bianchi identities for the supergeometry 
of section 2;
the investigation of six-dimensional supersymmetric backgrounds 
and projective superspace matter couplings as in the research
on 4D and 5D anti-de-Sitter supergeometries 
\cite{KT-M,Kuzenko:2008kw,Kuzenko:2008qw,Butter:2011kf,Butter:2012jj};
a more systematic classification of covariant projective superfields in six dimensions;
and the component reduction of the 6D projective action principle, for example along the lines of 
\cite{KT-M_5D,KT-M-normal}, which, within our formalism, is a first step towards the analysis of 
various supergravity-matter systems in components (see {\it e.\,g.} \cite{Butter:2012xg}).

Finally, we mention that new results on the 
construction of higher-derivative supergravity actions in six-dimensions have been obtained in
\cite{HD-6D}. It would be interesting to understand how classes of higher-derivative actions are 
constructed in six-dimensional projective superspace. 

\paragraph{Acknowledgements}
We are grateful to Sergei M.\,Kuzenko and Ulf Lindstr\"om for comments and collaboration at an 
early stage of 
this project, Daniel Butter and Joseph Novak for several comments, 
and C\'esar Arias H. for discussions of calculations 
related to the tensor multiplet. W{\sc dl}3 thanks the Department of Physics and Astronomy at 
Uppsala University, where part of this work was carried out, for their hospitality.
GT-M thanks the School of
Physics at the University of Western Australia and the Theory Unit at CERN for the kind
hospitality and support  during various stages of this work.
 He also thanks the Departments of Physics of Milano-Bicocca
University and Milano University for kind hospitality.
W{\sc dl}3 is supported by {\sc fondecyt} grant number 11100425 and {\sc dgid-unab} internal 
grant  DI-23-11/R.
The work of GT-M was supported by the European 
Commission, Marie Curie Intra-European Fellowships under contract 
No.~PIEF-GA-2009-236454.  GT-M is the recipient of an Australian Research Council's
Discovery Early Career Award (DECRA), project No.~DE120101498.

\appendix
\section{Six-dimensional Notation and Conventions}
\setcounter{equation}{0}
\label{appA}

Our six-dimensional superspace conventions are obtained by lifting the five-dimensional conventions established in references \cite{Kuzenko:2005sz,KT-M,KT-M_5D}.
 The procedure is to first define $\gamma_{{} a}:= - \Gamma_{{} a} C^{-1}$ and $\tilde \gamma_{{} a} = - C\Gamma_{{} a}$ for ${} a=0,\dots, 3;5$. Then we take $\gamma_6 = C^{-1}$ and $\tilde \gamma_6 = -C$.\footnote{
Keeping this procedure in mind, it is easier to verify certain statements using formulae from five dimensions. For example, since $\gamma_{{} a}^\mathrm{T} = (C^{-1})^\mathrm{T} \Gamma_{{} a}^\mathrm{T} = -C^{-1} C \Gamma_{{} a} C^{-1}=-\gamma_{{} a}$ for ${{} a}= 0,1,\dots,3;5$ we need inspect only $\gamma_6$ to conclude that these matrices are anti-symmetric.
} The relative sign has been chosen so that the six $8\times 8$ Dirac matrices satisfy
\begin{eqnarray}
\left\{ \Gamma_{{} a}, \Gamma_{{} b} \right\}  = -2 \eta_{{} ab} \mathbf 1
~,
\end{eqnarray}
with 
$a =0 ,\dots,3;5,6$
and
\bea
(\eta_{ab})=\rm{diag}(-1,1,1,1,1,1)~.
\eea
The overall sign is chosen so that, in terms of explicit indices, the formulae are
\begin{eqnarray}
(\gamma^a)_{\alpha \beta} = (\Gamma^a)_{\alpha \beta}
&,~~~&
(\tilde \gamma^a)^{\alpha \beta} = - (\Gamma^a)^{\alpha \beta}
~~~\mathrm{for}~ a=0,1,2,3;5\cr
(\gamma_6)_{\alpha \beta} = \varepsilon_{\alpha \beta} &,~~~&(\tilde \gamma_6)^{\alpha \beta}=-\varepsilon^{\alpha \beta}
~.
\end{eqnarray}
In terms of Pauli-type matrices, the Dirac matrices take the form
\begin{eqnarray}
\Gamma_{{} a} = \left(
\begin{array}{cc}
0& (\gamma_{{} a})_{{} \alpha {} \beta}\\ 
(\tilde \gamma_{{} a})^{{} \beta {} \alpha}&0
\end{array}
\right)
\end{eqnarray}
with ${} \alpha = 1,\dots,4$. 
We can give an explicit representation of $\g_a,\,\tilde{\g}_a$
in terms of the 4D Pauli matrices. In particular, denoting the 4D, SL$(2,\mathbb{C})$ spinor indices by
${\underline{\a}}=1,2$ and ${\dot{\underline{\a}}}=1,2$,
we use the representation
\begin{eqnarray}
\label{GammaMap1}
\gamma_a= \left(
\begin{array}{cc}
0&- (\sigma_a)_{{\underline{\alpha}}}{}^{\dot {\underline{\beta}}}
\\ 
(\tilde \sigma_a)^{\dot {\underline{\alpha}}}{}_{ {\underline{\beta}}}&0
\end{array}
\right)
~~~
\end{eqnarray}
for $a=0,\dots,3$ and 
\begin{eqnarray}
\label{GammaMap2}
\gamma_5 = \left(
\begin{array}{cc}
\ri \varepsilon_{{\underline{\alpha}}{\underline{ \beta}}}&0\\ 
0& 
\ri \varepsilon^{\dot {\underline{\alpha }}\dot {\underline{\beta}}}
\end{array}
\right)
~,~~~~
\gamma_6 = \left(
\begin{array}{cc}
\varepsilon_{{\underline{\alpha}} {\underline{\beta}}}&0\\ 
0& - \varepsilon^{\dot {\underline{\alpha}} \dot {\underline{\beta}}}
\end{array}
\right)
~~~
\end{eqnarray}
and similarly for matrices with upper indices.
They obey the Pauli-type algebra
\begin{eqnarray}
(\gamma^{{} a})_{{} \alpha {} \beta}( \tilde \gamma^{{} b})^{{} \beta {} \gamma}
	+(\gamma^{{} b})_{{} \alpha {} \beta}(\tilde \gamma^{{} a})^{{} \beta {} \gamma} 
	= -2 \eta^{{}ab} \delta_{{} \alpha}^{{} \gamma}~,
\cr
( \tilde \gamma^{{} a})^{{} \alpha {} \beta}(\gamma^{{} b})_{{} \beta {} \gamma}
	+( \tilde \gamma^{{} b})^{{} \alpha {} \beta}(\gamma^{{} a})_{{} \beta {} \gamma} 
	= -2 \eta^{ab} \delta^{{} \alpha}_{{} \gamma}~.
\end{eqnarray}
Due to our sign choices, the five-dimensional subalgebra agrees with that of references \cite{Kuzenko:2005sz, KT-M,KT-M_5D}.

Note that the six-dimensional Pauli-type matrices are antisymmetric
\begin{eqnarray}
(\gamma_{{} a})_{{} \alpha {} \beta}  = -(\gamma_{{} a})_{{} \beta {} \alpha}
~,
\end{eqnarray}
implying an isomorphism between the space of six-dimensional vectors and antisymmetric $4\times 4$ spin matrices
\begin{eqnarray}
V_{{} \a{} \b}:= (\gamma^{{} a})_{{}\a {}\b}V_{{} a}=-V_{{} \b {} \a}
&\Leftrightarrow& V_{{} a} = \tfrac14 (\tilde \gamma_{{} a})^{{} \alpha {} \beta} V_{{} \alpha {} \beta}
~.
\end{eqnarray}
The second relation is a consequence of the analysis below and equation (\ref{6Fierz2}) in particular.
Similarly, six-dimensional 2-forms are in one-to-one correspondence with traceless $4\times 4$ matrices and (anti-)self-dual 3-forms are in correspondence with symmetric rank-2 spin matrices with their indices (down) up as we now work out in detail. 

To begin, it is useful to define the normalized anti-symmetrized products of Pauli-type matrices 
\begin{align}
\begin{array}{lll}
\gamma_{ab} := \gamma_{[a}\tilde \gamma_{b]}:= \tfrac12\left( \gamma_a\tilde \gamma_b-\gamma_b\tilde \gamma_a\right)
~,
&~&
\gamma_{abc} := \gamma_{[a}\tilde \gamma_{b}\gamma_{c]}:= \tfrac1{3!}\left( \gamma_a\tilde \gamma_b\gamma_c \pm\mathrm{perm.}\right)
~,
\cr
\tilde \gamma_{ab} := \tilde \gamma_{[a}\gamma_{b]}:= \tfrac12\left( \tilde \gamma_a\gamma_b-\tilde \gamma_b\gamma_a\right)
~,
&~&\tilde \gamma_{abc} := \tilde \gamma_{[a}\gamma_{b}\tilde \gamma_{c]}:= \tfrac1{3!}\left( \tilde \gamma_a\gamma_b\tilde \gamma_c \pm \mathrm{perm.} \right)
~.
\end{array}
\end{align}
For example, this normalization implies
\begin{eqnarray}
{\gamma}^{ab} \gamma^c = {\gamma}^{abc} +2\eta^{c[a}\gamma^{b]} 
~,~~~
\tilde \gamma^c {\gamma}^{ab}  = \tilde {\gamma}^{abc} -2\eta^{c[a}\tilde \gamma^{b]}
~.
\end{eqnarray}
On the other hand, a more commonly used convention regarding the 2-form matrix is as the spinor representation (\ref{LorentzSpin}) of the Lorentz generator $M_{ab}$ which is related by
\begin{eqnarray}
\label{SpinNorm}
(\Sigma^{ab})_\alpha{}^\beta = - \tfrac12 (\gamma^{ab})_\alpha{}^\beta
~.
\end{eqnarray}
In terms of these matrices, we define
\begin{eqnarray}
F_{{} \a}{}^{{} \b}:= \tfrac12 ({\Sigma}^{ab})_{{}\a}{}^{{}\b}F_{ab}
&\Rightarrow&  F_{ab}= - ({\Sigma}_{ab})_{{}\b}{}^{{}\a} F_{{} \a}{}^{{} \b}
~.
\end{eqnarray}
The second relation is a consequence of (\ref{Fierz2form}) which follows from the
analysis below.
Using the second type of matrix, we can construct $\tilde{F}^{{} \a}{}_{{} \b}:= (\tilde{{\gamma}}^{ab})^{{}\a}{}_{{}\b}F_{ab}$ but $(\tilde {\gamma}^{ab})^{{} \alpha}{}_{{} \beta} = - ({\gamma}^{ab})_{{} \beta}{}^{{} \alpha}$ so that this second matrix is not essentially new. 

Finally, the third-rank antisymmetric tensors can be separated into (anti-)self-dual parts which are then in one-to-one correspondence with symmetric $4\times4$ matrices. To see how this works in detail, we must first establish some Fierz identities. There is a completeness relation
\begin{eqnarray}
\label{6Fierz1}
\tfrac12 (\gamma^{{} a})_{{} \alpha {} \beta} (\gamma_{{} a})_{{} \gamma {} \delta} = \varepsilon_{{} \alpha {} \beta {} \gamma {} \delta}
~.
\end{eqnarray}
Contraction with $\varepsilon^{{} \gamma^\prime {} \delta^\prime {} \gamma {} \delta}$ implies the completeness relation
\begin{eqnarray}
\label{6Fierz2}
\tfrac12 (\gamma^{{} a})_{{} \alpha {} \beta} (\tilde \gamma_{{} a})^{{} \gamma {} \delta} = \delta_{{} \alpha}^{ {} \gamma} \delta_{ {} \beta}^{ {} \delta} -\delta_{{} \beta}^{{} \gamma} \delta_{ {} \alpha}^{ {} \delta}
\end{eqnarray}
and that
\begin{eqnarray}
\tfrac12 \varepsilon^{{} \alpha {} \beta {} \gamma {} \delta} (\gamma_{{} a})_{{} \gamma {} \delta} =  (\tilde \gamma_{{} a})^{{} \alpha {} \beta}
&\Rightarrow&
(\gamma_{{} a})_{{} \alpha {} \beta} =\tfrac12 \varepsilon_{{} \alpha {} \beta {} \gamma {} \delta}   (\tilde \gamma_{{} a})^{{} \gamma {} \delta}
~.
\end{eqnarray}
Contraction of (\ref{6Fierz2}) with itself gives
\begin{eqnarray}
\label{Fierz2form}
\tfrac 14 (\tilde {\gamma}^{ab})^{{} \alpha}{}_{{} \beta}({\gamma}_{ab})_{{} \gamma}{}^{{} \delta} = -\tfrac12 \delta_{{} \beta}^{{} \alpha}\delta_{{} \gamma}^{{} \delta} +2 \delta_{{} \beta}^{{} \delta}\delta_{{} \gamma}^{{} \alpha}
~.
\end{eqnarray}
Another contraction with (\ref{6Fierz2}) gives
\begin{eqnarray}
\label{Fierz3form}
(\tilde {\gamma}^{abc})^{{} \alpha {} \beta} ({\gamma}_{abc})_{{} \gamma {} \delta} =24(\delta^{{} \alpha}_{{} \gamma}\delta^{{} \beta}_{{} \delta} + \delta^{{} \alpha}_{{} \delta}\delta^{{} \beta}_{{} \gamma})
~,
\end{eqnarray}
while contraction with (\ref{6Fierz1}) shows that 
\begin{eqnarray}
({\gamma}^{abc})_{{} \alpha {} \beta} ({\gamma}_{abc})_{{} \gamma {} \delta} = 0 
	&\mathrm{and}& 
		(\tilde {\gamma}^{abc})^{{} \alpha {} \beta}( \tilde {\gamma}_{abc})^{{} \gamma {} \delta} = 0~.
\end{eqnarray}
Thus we see that $\tilde {\gamma}^{abc}$ and ${\gamma}^{abc}$ correspond to (anti-)self-dual 3-forms.
To show that (${\gamma}^{abc}$) $\tilde {\gamma}^{abc}$ is (anti-)self-dual, one can use the identities 
\begin{eqnarray}
\gamma^0\tilde \gamma^1\gamma^2\tilde \gamma^3\gamma^5\tilde \gamma^6 = -\mathbf 1
	~~~\mathrm{and}~~~
\tilde \gamma^0\gamma^1\tilde \gamma^2\gamma^3\tilde \gamma^5 \gamma^6 = +\mathbf 1
~,
\end{eqnarray}
to conclude that, for example, $\gamma^{012} = - \epsilon^{012}{}_{345} \gamma^{345}$ whereas $\tilde \gamma^{012} =  \epsilon^{012}{}_{345} \tilde \gamma^{345}$. Therefore, to conform to the accepted conventions on (anti-)self-duality, we normalize $\epsilon^{012356}=1$.

From the trace relation on the 3-forms
\begin{eqnarray}
\mathrm {tr} (\tilde \gamma_{abc}\gamma^{def}) = 
	4! \left( \delta_{[a}^d\delta_b^e\delta_{c]}^f +\tfrac1{3!} \epsilon_{abc}{}^{def} \right) 
~,
\end{eqnarray}
it follows that the (anti-)self-dual parts of a 3-form $N$ satisfy
\begin{eqnarray}
N_{\alpha \beta}:= \tfrac1{3!} N_{abc} (\gamma^{abc})_{\alpha \beta} 
	~~&\Rightarrow&~~  N^{(+)}_{abc}=\tfrac18 N_{\alpha \beta} (\tilde\gamma_{abc})^{\alpha \beta}
~,	\cr
N^{\alpha \beta}:= \tfrac1{3!} N_{abc} (\tilde \gamma^{abc})^{\alpha \beta} 
	~~&\Rightarrow&~~  N^{(-)}_{abc}=\tfrac18 N^{\alpha \beta} (\gamma_{abc})_{\alpha \beta} 
~.
\end{eqnarray}
In six dimensions, Hodge duality on 3-forms is an involution of order 2:
\begin{eqnarray}
\tfrac 1{3!} \epsilon_{abcrst}\epsilon^{defrst} = - 3! \delta_{[a}^d\delta_b^e \delta_{c]}^f
~.
\end{eqnarray}

Following  \cite{Kuzenko:2005sz}, the components of the SU$^*(4)$ 
spinor and its complex conjugate 
\begin{eqnarray}
(\Psi_{{} \alpha}) = \left(
\begin{array}{c}
\psi_{\underline{\alpha}}\\
\bar \phi^{\dot {\underline{\alpha}}}
\end{array}
\right)
~~~\mathrm{and}~~~
(\Psi^*_{{}\bar  {{} \alpha}} )= \left(
\begin{array}{c}
\bar \psi_{\dot {\underline{\alpha}}}\\
\phi^{{\underline{\alpha}}}
\end{array}
\right)
\end{eqnarray}
are defined in terms of 4D SL($2,\mathbb{C}$)
spinors.
Introducing the unitary matrix 
\bea
\left(B_{{} \alpha}{}^{{}\bar  {{} \beta}}\right)= \left(
\begin{array}{cc}
0 & \varepsilon_{{\underline{\alpha}}{\underline{ \beta}}}\\
-\varepsilon^{\dot {\underline{\alpha}} \dot {\underline{\beta}}}&0
\end{array}
\right)
&\Rightarrow&
\left(B^{\mathrm T\, {}\bar  {{} \beta}}{}_{{} \alpha}\right)= \left(
\begin{array}{cc}
0 & \varepsilon^{ \dot {\underline{\beta}} \dot {\underline{\alpha}}} \\
-\varepsilon_{ {\underline{\beta}}{\underline{\alpha}}}&0
\end{array}
\right)
~,
\eea
it can be checked explicitly that, using the representation defined by (\ref{GammaMap1}) and (\ref{GammaMap2}),
\begin{eqnarray}
B_{{} \alpha}{}^{{}\bar  {{} \alpha} } B_{{} \beta}{}^{{}\bar  {{} \beta} } (\gamma^*_{{} a})_{{}\bar  {{} \alpha} {}\bar  {{} \beta} } = (\gamma_{{} a})_{{} \alpha {} \beta}
~.
\end{eqnarray}
We may, therefore, define the covariant conjugate\footnote{The signs have been chosen so that the covariant conjugate reduces in five dimensions to the (transpose of the) Dirac conjugate
in the conventions of \cite{Kuzenko:2005sz}. 
}
\begin{eqnarray}
\overline{ (\Psi_{{} \alpha})} :=( B_{{} \alpha}{}^{{}\bar  {{} \alpha}} \, \Psi^*_{{}\bar  {{} \alpha}})
	= \left(
	\begin{array}{c}
\phi_{\underline{\alpha}}\\
-\bar \psi^{\dot {\underline{\alpha}}}
\end{array}
\right)~.
\end{eqnarray}
The complex conjugate $\left(B^*_{{}\bar  {{} \alpha}}{}^{{} \beta}\right)$ satisfies
\begin{eqnarray}
BB^*= -\mathbf 1 = B^*B
~.
\end{eqnarray}
This implies that performing the conjugation twice,
\begin{eqnarray}
\overline {\overline \Psi} = \overline {( B \Psi^*)} = B(B^* \Psi)=-\Psi
~.
\end{eqnarray}
Finally, we define the doublet
\begin{eqnarray}
(\Psi_{\alpha}^{ i})
	~~~\mathrm{such~that}~~~
	\Psi_{ \alpha}^{ \underline 1}= \Psi_\alpha 
	~~~\mathrm{and}~~~
	\Psi_{\alpha}^{ \underline 2}= \bar \Psi_\alpha
	~.
\end{eqnarray}
This combination satisfies the SU(2)-Majorana-Weyl reality condition
\begin{eqnarray}
\overline{(\Psi_\a^i)}=
B_{{} \alpha}{}^{{}\bar  {{} \alpha}} ((\Psi^i)^*)_{{}\bar  {{} \alpha}} = \Psi_{{} \alpha}{}_i
~,
\end{eqnarray}
where we used the normalization $\varepsilon_{\underline 2 \underline 1}=1$. It follows from this that 
\begin{eqnarray}
\overline{\overline {\Psi^i}} = \Psi^i
~.
\end{eqnarray}
Finally, the following conjugation relation holds
\begin{eqnarray}
\overline {(\mathcal D_{{} \alpha}^{ i} \Phi)}= - (-1)^{|\Phi|} \mathcal D_{{} \alpha i} \bar \Phi
~.
\end{eqnarray}

\section{On the Analytic Projector}
\setcounter{equation}{0}
\label{AnalyticProjectorAppendix}
In this appendix we derive the results stated at the end of section \ref{projSuper}
about the analytic projection operator $\D^{(4)}$ equations (\ref{anapro}) and (\ref{anapro-2}).

First of all we, want to prove that, given a general scalar weight-$n$ isotwistor superfield $U^{(n)}$, the superfield $Q^{(n+4)}:=\D^{(4)}U^{(n)}$ satisfies $\cD_\a^{(1)}Q^{(n+4)}=0$.
The derivation goes along the same lines of the 5D case of 
\cite{KT-M,KT-M_5D,KT-M_5Dconf}.\footnote{The analysis in 
this section generalize and in principle can be used to derive the analytic projection operator
for the conformal supergravity geometry of \cite{KT-M_5Dconf} in the case where the 5D
$C^{ij}_{\hat{a}}$ torsion component is nonzero. Such case was not presented in 
\cite{KT-M_5Dconf}.}

The starting point is to observe that by construction
\bea
\cD^{(1)}_{[\a}\cD^{(1)}_\b\cD^{(1)}_\g\cD^{(1)}_\d\cD^{(1)}_{\r]}=0
~,
\eea
which implies
\bea
\cD_\a^{(1)}\cD^{(4)}
&=&
-\tfrac{1}{480}\ve^{\b\g\d\r}\Big{(}
4\{\cD^{(1)}_{\a},\cD^{(1)}_\b\}\cD^{(1)}_\g\cD^{(1)}_\d\cD^{(1)}_{\r}
+3\cD^{(1)}_{\g}\{\cD^{(1)}_\a,\cD^{(1)}_\b\}\cD^{(1)}_\d\cD^{(1)}_{\r}
\non\\
&&
+2\cD^{(1)}_\g\cD^{(1)}_\d\{\cD^{(1)}_{\a},\cD^{(1)}_\b\}\cD^{(1)}_{\r}
+\cD^{(1)}_\g\cD^{(1)}_\d\cD^{(1)}_{\r}\{\cD^{(1)}_{\a},\cD^{(1)}_\b\}
\Big{)}
~.
\eea
Then, one applies the previous equation on the superfield $U^{(n)}$ and 
compute the anti-commutators. 
Since $[J^{(2)},\cD_\a^{(1)}]=J^{(2)}U^{(n)}=0$, the SU(2) part of the anti-commutator
algebra (\ref{D1D1}) does not contribute at all in the computation.
On the other hand, from the Lorentz part of (\ref{D1D1}) we need to systematically move
to the right the Lorentz generator by using 
$[ M_{{} \alpha}{}^{ {} \beta}, \mathcal D_{{} \gamma k} ]=\big( \tfrac14\delta_{{} \alpha}^{{} \beta} \mathcal D_{{} \gamma k} - \delta_{{} \gamma}^{{} \beta} \cD_{\a k}\big)$
 to then hit $U^{(n)}$ and use $M_\a{}^\b U^{(n)}=0$.
 Moreover, one needs to elaborate on relations involving multiple $\cD^{(1)}$ derivatives of 
 $C_{\a\b}^{(2)}$.
It is easy to prove that
\bea
\{ \cD^{(1)}_\a, \cD^{(1)}_\b \}C_{\g\d}^{(2)}&=&0
~~~~~~
\Leftrightarrow
~~~~~~
\cD^{(1)}_\a\cD^{(1)}_\b C_{\g\d}^{(2)}=
-\cD^{(1)}_\b\cD^{(1)}_\a C_{\g\d}^{(2)}
=\cD^{(2)}_{\a\b} C_{\g\d}^{(2)}
~.
\label{DDC++++}
\eea
On the other hand, due to (\ref{solN}),
we know that
\bea
\cD_\a^{(1)}C_{\b\g}^{(2)}=\cD_{[\a}^{(1)}C_{\b\g]}^{(2)}~,~~~~~~
\cD_\a^{(1)}C_{\b\g}^{(2)}=2\ve_{\a\b\g\d}\cC^{(3)}{}^\d
~.
\eea
It is then clear that the following equation holds
\bea
\cD_\a^{(1)}\cD_\b^{(1)}C_{\g\d}^{(2)}&=&
\cD_{[\a}^{(1)}\cD_{\b}^{(1)}C_{\g\d]}^{(2)}
=
\tfrac{1}{12}\ve_{\a\b\g\d}(\cD_{\r\t}^{(2)}C^{(2)}{}^{\r\t})
~.
\eea
One can also derive this further result 
\bea
\cD^{(1)}_{\a}\cD^{(2)}_{\b\g} C_{\d\r}^{(2)}
&=&
-12\ri \cD^{(1)}_{\a}C^{(2)}_{[\b\g} C_{\d\r]}^{(2)}
~.
\eea
At this point, after some algebra, one can obtain
\bea
\cD_\a^{(1)}\cD^{(4)}U^{(n)}
&=&
\ve^{\b\g\d\r}\Big{(}
\tfrac{5\ri}{12} C^{(2)}_{\b\g}\cD^{(1)}_{\a}\cD^{(2)}_{\d\r}
-\tfrac{5\ri}{36}(\cD^{(1)}_{\b}C^{(2)}_{\g\d})\cD^{(1)}_{\a}\cD^{(1)}_{\r}
\non\\
&&
+\tfrac{\ri}{48}(\cD^{(2)}_{\b\g}C^{(2)}_{\d\r})\cD^{(1)}_{\a}
- \tfrac{3}{2}C^{(2)}_{\b\g}C^{(2)}_{\d\r}\cD^{(1)}_{\a}
\Big{)}
U^{(n)}
\label{DDDDD+++++}
\eea
It is then easy to get the following results
\bsubeq
\bea
\cD^{(1)}_\a\big(\ve^{\b\g\d\r}C^{(2)}_{\b\g}\cD^{(2)}_{\d\r}U^{(n)}\big)
&=&
\ve^{\b\g\d\r}\Big(
 C^{(2)}_{\b\g}\cD^{(1)}_\a\cD^{(2)}_{\d\r}
+\tfrac{2}{3}(\cD^{(1)}_\b C^{(2)}_{\g\d})\cD^{(1)}_\a\cD^{(1)}_{\r}
\Big)U^{(n)}
~,~~~~~~
\\
\cD^{(1)}_\a\big(\ve^{\b\g\d\r}(\cD^{(1)}_\b C^{(2)}_{\g\d})\cD^{(1)}_{\r}U^{(n)}\big)
&=&
\ve^{\b\g\d\r}\Big(
-(\cD^{(1)}_\b C^{(2)}_{\g\d})\cD^{(1)}_\a\cD^{(1)}_{\r}
\non\\
&&~~~~~~~~~
-\tfrac{1}{4}(\cD^{(2)}_{\b\g} C^{(2)}_{\d\r})\cD^{(1)}_{\a}
\Big)U^{(n)}
~.
\\
\cD^{(1)}_\a\big(\ve^{\b\g\d\r}(\cD^{(1)}_\b \cD^{(1)}_{\g}C^{(2)}_{\d\r})U^{(n)}\big)
&=&
\ve^{\b\g\d\r}\Big(
(\cD^{(2)}_{\b\g} C^{(2)}_{\d\r})\cD^{(1)}_\a
-12\ri\big(\cD^{(1)}_{\a}C^{(2)}_{\b\g}C^{(2)}_{\d\r}\big)
\Big)U^{(n)}
~,~~~~~~~~~~~
\\
\cD^{(1)}_\a\big(\ve^{\b\g\d\r}(C^{(2)}_{\b\g}C^{(2)}_{\d\r})U^{(n)}\big)
&=&
\ve^{\b\g\d\r}\Big(
(C^{(2)}_{\b\g}C^{(2)}_{\d\r})\cD^{(1)}_\a
+\big(\cD^{(1)}_\a C^{(2)}_{\b\g}C^{(2)}_{\d\r}\big)
\Big)U^{(n)}
~.
\label{ffffff}
\eea
\esubeq
By using the equations (\ref{DDDDD+++++})--(\ref{ffffff})
one can then observe that the combination of operators in the analytic projection operator (\ref{anapro})
is such that
\bea
\cD_\a^{(1)}\D^{(4)}U^{(n)}
&=&
0
~.
\eea

As a next step we want to compute the super-Weyl transformations of
the superfield $ Q^{(n+4)} = \D^{(4)} U^{(n)}$ supposing that 
$U^{(n)}$ transforms homogeneously $\d_\s U^{(n)}=w\s U^{(n)}$.
To do that, we need some straightforward intermediate results.
In particular, we have
\bsubeq
\bea
\d_\s\cD_{\a}^{(1)}&=&\tfrac12\s\cD_{\a}^{(1)}
-2(\cD_{\b}^{(1)}\s)M_{\a}{}^\b
+4(\cD_{\a}^{(1)}\s)J^{(0)}
-4(\cD_{\a}^{(-1)}\s)J^{(2)}
~,
\label{sWD1}
\\
\d_\s C_{\a\b}^{(2)}&=&
\s C_{\a\b}^{(2)}
+\tfrac {\ri}2(\cD^{(2)}_{\a\b}\s)
~,
\label{sWC2}
\eea
\esubeq
where
\bsubeq
\bea
J^{(0)}&:=&\frac{v_iu_j}{(v,u)} J^{ij}~,~~~
{[}J^{(0)},\cD_\g^{(1)}{]}=-\frac{1}{2}\cD_\g^{(1)}~,~~~
J^{(0)}U^{(n)}=-\frac{n}{2}U^{(n)}~,
\\
\cD_\a^{(-1)}&:=&\frac{u_i}{(v,u)} \cD_\a^i~.
\eea
\esubeq
The $(\cD_\a^{(-1)}\s)$ term in (\ref{sWD1})
does not actually enter into this computation since $J^{(2)}$
commutes with the $\cD_\a^{(1)}$ derivatives and $J^{(2)}U^{(n)}=0$.
Defining,
\bsubeq
\bea
&\s^{(1)}_\a:=\cD_\a^{(1)}\s~,~~~
\s^{(2)}_{\a\b}:=\cD_\a^{(1)}\cD_\b^{(1)}\s~,~~~
\\
&
\s^{(3)}_{\a\b\g}:=\cD_{[\a}^{(1)}\cD_\b^{(1)}\cD_{\g]}^{(1)}\s~,~~~
\s^{(4)}_{\a\b\g\d}:=\cD_{[\a}^{(1)}\cD_\b^{(1)}\cD_\g^{(1)}\cD_{\d]}^{(1)}\s
~,~~~~~~~~~
\eea
\esubeq
we obtain other necessary intermediate results
\bsubeq
\bea
\d_\s\cC^{(3)\a}&=&
\tfrac32\s\cC^{(3)\a}
+\tfrac{4}{3}\s_{\b}^{(1)} C^{(2)\a\b}
-\tfrac{\ri}{24}\ve^{\a\b\g\d}\s_{\b\g\d}^{(3)}
~,~~~~~~
\\
\d_\s\Big(
\cD^{(2)}_{\g\d}C^{(2)\g\d}
\Big)&=&
\Big(
2\s\big(\cD^{(2)}_{\g\d}C^{(2)\g\d}\big)
+120\s_{\a}^{(1)}\cC^{(3)\a}
-8\s^{(2)}_{\g\d}C^{(2)\g\d}
+\tfrac{\ri}{4}\ve^{\a\b\g\d}\s^{(4)}_{\a\b\g\d}
\Big)
~.~~~~~~~~~~
\eea
\esubeq
Finally, after some algebra, one can obtain the following equation
\bea
\d_\s\D^{(4)}U^{(n)}&=&
(w+2)\s\D^{(4)}U^{(n)}
\non\\
&&
+(w-2n-6)\Big(
-\tfrac{1}{24}\ve^{\a\b\g\d}\s_{\a}^{(1)}\cD_\b^{(1)}\cD_\g^{(1)}\cD_\d^{(1)}U^{(n)}
-\tfrac{1}{16}\ve^{\a\b\g\d}\s_{\a\b}^{(2)}\cD_{\g\d}^{(2)} U^{(n)}
\non\\
&&~~~
-\tfrac{1}{24}\ve^{\a\b\g\d}\s_{\a\b\g}^{(3)}\cD_\d^{(1)} U^{(n)}
-\tfrac{1}{96}\ve^{\a\b\g\d}\s_{\a\b\g\d}^{(4)} U^{(n)}
-\tfrac{5\ri}{3}\s_{\a}^{(1)}C^{(2)\a\b}\cD^{(1)}_{\b}U^{(n)}
\non\\
&&~~~
-\tfrac{5\ri}{6}\s_{\a\b}^{(2)}C^{(2)\a\b}U^{(n)}
+5\ri\,\s^{(1)}_{\a}\cC^{(3)\a}U^{(n)}
\Big)
~.
\eea
It is clear that by choosing 
\bea
w=2(n+3)~,
\eea
the weight-$(n+4)$ projective superfield $Q^{(n+4)} = \D^{(4)} U^{(n)}$
transforms homogenously, $\d_\s  Q^{(n+4)} =2(n+4)\s Q^{(n+4)}$, 
in agreement with equation (\ref{super-Weyl-Qn}).

To conclude this appendix, we point out that, in the case $n=0$, all the previous results 
are exactly the same if instead of a weight-0 isotwistor superfield $U^{(0)}$ 
one considers a $v$-independent superfield $P(z)$ such that $\d_\s P=6\s P$.
Then, the superfield $\cP^{(4)}:=\D^{(4)}P$ is a weight-$4$ projective superfield.
To convince oneself of this,
one has only to notice that for both $U^{(0)}$ and $P$ the conformal 
weight is $6$ and that $J^{(2)}U^{(0)}=J^{(0)}U^{(0)}=J^{(2)}P=J^{(0)}P=0$ holds.

\footnotesize{

}

\end{document}